\documentclass[preprint,12pt]{elsarticle}

\usepackage{amssymb}
\usepackage{amsmath}
\usepackage{algorithm}
\usepackage{algpseudocode}

\journal{Computers & Mathematics with Applications}

\begin{document}

\begin{frontmatter}



\title{Sparse and low-rank kinetic distribution estimation} 

\author[label1]{Georgii Oblapenko}
\author[label1]{Lambert Theisen}
\author[label2]{Rostislav-Paul Wilhelm}
\author[label3]{Michael Herty}
\author[label1]{Manuel Torrilhon}
\affiliation[label1]{organization={Applied and Computational Mathematics, RWTH Aachen},
addressline={Schinkelstraße 2},
city={Aachen},
postcode={52062},
country={Germany}}

\affiliation[label2]{organization={Centre for mathematical Plasma Astrophysics, KU Leuven},
addressline={Celestijnenlaan 200b},
city={Leuven},
postcode={3001},
country={Belgium}}

\affiliation[label3]{organization={Institut für Geometrie und praktische Mathematik, RWTH Aachen},
addressline={Templergraben 55},
city={Aachen},
postcode={52062},
country={Germany}}

\begin{abstract}
In this paper, we consider methods that allow for memory-efficient storage of high-dimensional
distributions and retain certain key features thereof, specifically
in a kinetic theory context. We propose an extension to the entropic quadrature
method that allows for enforcing sparsity, and propose a new low-rank
decomposition approach that ensures preservation of moment information. The
methods are applied to several model kinetic distributions, as well as to
distributions obtained from high-resolution kinetic simulations of the
Vlasov--Maxwell system.
\end{abstract}




\end{frontmatter}



\section{Introduction}
\label{sec:intro}
Kinetic equations are used to describe a wide range of real-life phenomena involving movement and cross-interaction of particles or particle-like agents, such
as rarefied gas flows~\cite{cercignani2000rarefied}, plasma flows~\cite{chen2015introduction}, radiation transport~\cite{modest2021radiative}, social dynamics~\cite{boscheri2021modeling},
or traffic flow~\cite{herty2020bgk}. Kinetic equations describe the spatio-temporal evolution of a probability distribution that describes the distribution of the particle population in
phase (parameter) space. A common feature of kinetic equations is their high dimensionality, which precludes the use of direct solution methods~\cite{dimarco2014numerical} for large-scale problems, and complicates
storage of their solutions.
To remedy this, approaches such as dynamical low-rank methods have been developed~\cite{koch2007dynamical,bachmayr2023low,einkemmer2025review}. These utilize a tensor decomposition of an ansatz
for the distribution function, and consider evolution equations for the expansion coefficients, thereby reducing the dimensionality of the system of equations. Recent developments in this area include conservative dynamical low-rank (DLR)
methods for the Vlasov~\cite{allmann2022parallel,guo2024conservative}
and Boltzmann equations~\cite{chikitkin2021numerical,hu2022adaptive,oblapenko2023use,dektor2025interpolatory,einkemmer2025asymptotic} in plasma and gas dynamics, respectively.
There has been also some research towards conservative DLR as well as step-and-truncate (SAT) approaches, focussing on the conservation of lower order moments~\cite{einkemmer2025asymptotic,einkemmer2025review,guo2024conservative,el2024krylov,coughlin2024robust}; these efforts are summarized in the recent review article~\cite{einkemmer2025review}.

Direct imposition of sparsity on a kinetic distribution function has been less studied, with our recent work~\cite{oblapenko2026sparse} investigating the possibility of introducing sparsity by using a non-fixed ansatz to
 represent the distribution, and then minimizing distance between the ansatz functions, thus effectively reducing their
 number.
 Although the approach allows for a significant degree of flexibility and avoids the curse of dimensionality usually
 associated with multi-dimensional problems, it requires minimization of a non-convex non-differentiable function.

 In the present work, we focus on two related problems: 1) recovering an unknown distribution function from known moments, i.e.\ the multi-dimensional Hamburger moment problem~\cite{hamburger1944hermitian}, whilst
 simultaneously enforcing sparsity or low-rank structure on the fixed ansatz used to represent the distribution, and 2) obtaining a sparse or low-rank representation of a given distribution whilst preserving the moments thereof.

Although the present work is mostly concerned with reconstruction
of underlying distribution functions without a specific focus on particular governing kinetic equations, the examples used for evaluation of the performance of the proposed algorithms
come from the Boltzmann and Vlasov--Maxwell equations. The paper is structured as follows: in Section~\ref{sec:governing-eqns}, we introduce the governing equations, the distribution functions and moments
thereof. In Section~\ref{sec:df-estimation} we discuss the ansatz used to represent the distribution function, and the computation of moments via this ansatz. We then propose an algorithm for
sparse estimation of distribution functions in Section~\ref{sec:speq}, and an algorithm for low-rank estimation of distribution functions in Section~\ref{sec:mclr}. We present numerical results in Section~\ref{sec:numres}, and finally provide conclusions in Section~\ref{sec:conclusions}.

\section{Governing equations}\label{sec:governing-eqns}
We consider the kinetic equation given by
\begin{equation}
    \partial_t f + \mathbf{v} \cdot \nabla_{\mathbf{x}} f + \frac{\mathbf{F}}{m} \cdot \nabla_{\mathbf{v}} f = Q(f,f),\label{eqn:kinetic}
\end{equation}
where $f(\mathbf{v},\mathbf{x},t): \mathbb{R}^{n_v} \times \mathbb{R}^{n_x} \times \mathbb{R}_{+} \to \mathbb{R}_{\geq 0 }$ is the distribution function
giving the expectation number of particles having a velocity $\mathbf{v} \in \mathbb{R}^{n_v}$ in a spatial location $\mathbf{x} \in \mathbb{R}^{n_x}$ at time $t$. $\mathbf{F}$ is an external force
acting on the particles, $m$ is the mass of the particles, and $Q(f,f)$ is the ``collision operator'' describing the interaction of particles. The terms on the left-hand
side correspond to advection and acceleration of the particles.
One particular case of the kinetic equation is the Boltzmann equation for a monatomic uncharged gas, in which case $\mathbf{F} = 0$, and $Q$ is given as (omitting
$\mathbf{x}$ and $t$ to simplify notation)
\begin{equation}
    Q(f,f) = \int_{\mathbb{R}^3} \int_{S^2}
(f(\mathbf{v}_1') f(\mathbf{v}') - f(\mathbf{v}_1) f(\mathbf{v})) \, |\mathbf{v} - \mathbf{v}_1| \, d\sigma \, d\mathbf{v}_1,
\end{equation}
where $\mathbf{v}_1$ is the velocity of the collision partners, $\sigma$ is the scattering angle parameterized on the unit sphere, and the primed
velocities $\mathbf{v}_1'$, $\mathbf{v}'$ denote the post-collisional velocities found from binary collision mechanics. The Boltzmann equation describes the behaviour of a non-dense gas
in a variety of rarefaction regimes.

Another set of particular cases are the collisionless Vlasov--Poisson and Vlasov--Maxwell equations: in both cases $Q(f,f)=0$, and $\mathbf{F} = q\mathbf{E}$ or
$\mathbf{F} = q\left(\mathbf{E} + \mathbf{v} \times \mathbf{B}\right)$ in the case of Vlasov--Poisson and Vlasov--Maxwell equations, respectively.
These equations describe the behaviour of a collisionless plasma with applied electric and magnetic fields $\mathbf{E}$ and $\mathbf{B}$, correspondingly; $q$ denotes the charge of the particles.
For a self-consistent description, the kinetic equation is coupled with the Poisson electrostatic or Maxwell electromagnetic equations, with the charge and current density computed via $f$.


As mentioned in the introduction, a uniting theme for kinetic equations is their high dimensionality, which leads to a high computational cost when attempting to directly solve~(\ref{eqn:kinetic}).
Storing the high-dimensional
solutions is also oftentimes problematic.
Assuming a tensor-product ansatz for $f$, we can write:
\begin{equation}
    f(\mathbf{v}) = \sum f_{k_1,\ldots,k_{n_v}} \psi_{k_1}(v_1) \cdot \ldots \psi_{k_{n_v}}(v_{n_v}),
\end{equation}
where $\psi_{k_i}$, $i=1,\ldots,n_v$ are fixed basis functions, and $f_{k_1,\ldots,k_{n_v}}$ the expansion coefficients,
indexed by an $n_v$-dimension multi-index $(k_1,\ldots,k_{n_v})$. The components of $\mathbf{v}$ are the scalar values~$v_1,\ldots,v_{n_v}$.
Low-rank methods reduce the storage requirements for $f$ by assuming a decomposition of the tensor of coefficients $f_{k_1,\ldots,k_{n_v}}$, for example, using the Tucker~\cite{koldaTensorDecompositionsApplications2009} ansatz:
\begin{equation}
    f_{k_1,\ldots,k_{n_v}} \approx \sum_{\alpha_1=1}^{r_1}\ldots\sum_{\alpha_{n_v}=1}^{r_{n_v}}
    C_{\alpha_1 \ldots \alpha_{n_v}} U^{(1)}_{k_1\alpha_1} \ldots U^{(k_{n_v})}_{k_{n_v}\alpha_{n_v}},\label{eq:low-rank-ansatz}
\end{equation}
where $C$ is a core tensor of size $r_1 \times \ldots \times r_{n_v}$, $r_1,\ldots,r_{n_v}$ are the ranks of the decomposition, and $U^{(i)}$ are vectors of length $n_{k_i}$.
Depending on the specific problem and time integration scheme used, one can operate on the coefficients of the low-rank representation~(\ref{eq:low-rank-ansatz}) of $f$ directly~\cite{koch2007dynamical},
without having to reconstruct the full-dimensional $f$.

Another approach to tackling the high dimensionality of~(\ref{eqn:kinetic}) is projection onto a set of basis functions $\phi_{k_1,\ldots,k_{n_v}}(\mathbf{v})$ in $\mathbb{R}^{n_v}$ and subsequently integrating over $\mathbf{v}$,
leading to a system of PDEs for the projections. In
case the basis functions are chosen as multi-variate monomials, i.e.
\begin{equation}
    \phi_{k_1,\ldots,k_{n_v}}(\mathbf{v}) = v_1^{k_1} \cdot \ldots \cdot v_{n_v}^{k_{n_v}},
\end{equation}
a set of \textit{moment equations} is obtained, which reads (omitting the acceleration and interaction terms for simplicity):
\begin{equation}
    \frac{\partial M_{k_1,\ldots,k_{n_v}}}{\partial t} + \nabla \cdot \mathcal{F}_{k_1,\ldots,k_{n_v}} = 0,\:k_l \geq 0,\: l=1,\ldots,n_v,\label{eq:transport}
\end{equation}
where the moments are defined as
\begin{equation}
    M_{k_1,\ldots,k_{n_v}}(\mathbf{x},t) = \int  f(\mathbf{v},\mathbf{x},t)\prod_{l=1}^{n_v} v_l^{k_l} \mathrm{d} \mathbf{v} \label{eq:moment-def},
\end{equation}
and the flux is given by
\begin{equation}
    \mathcal{F}_{k_1,\ldots,k_{n_v}} = \left(M_{k_1+1,k_2,\ldots,k_{n_v-1},k_{n_v}},M_{k_1,k_2+1,\ldots,k_{n_v-1},k_{n_v}},\ldots,M_{k_1,k_2,\ldots,k_{n_v-1},k_{n_v}+1}\right).
\end{equation}
The lower-order moments correspond to usual macroscopic quantities such as density, velocity, and energy.
It can be seen that the transport equations for a moment of given \textit{total order $m=\sum_l k_l$} involve moments of total order $m+1$, thus leading to an infinite hierarchy of moment equations.
Therefore, a procedure is required to compute a higher-order moment only on the basis of the knowledge of all lower-order moments. This can be carried out by estimating the unknown underlying distribution function
$f(\mathbf{v},\mathbf{x},t)$, thus in effect solving the multi-dimensional Hamburger moment problem~\cite{hamburger1944hermitian,Shohat1945problem,Aheizer1962,schmuedgen2017moment}, and computing the required
higher-order moments from it.

Although at first glance flux closures in moment equations and dynamical low-rank methods are not closely related, we re-formulate our research question as follows: is it possible, given knowledge about
moments of the distribution function (and, potentially, the underlying distribution itself as well), to reconstruct an approximated distribution that can be efficiently stored? In the next section, we
develop methods for such reconstructions.

\section{Distribution function estimation}\label{sec:df-estimation}
Since the application examples in the present work come from rarefied gas and plasma dynamics, we consider only a three-dimensional velocity space ($n_v=3$), and will use notation for 3-dimensional
tensors only with explicit naming of the directions $x$, $y$, and $z$ instead of their enumeration; however, the methods presented here are easily extended to higher dimensions.

We also decouple the velocity $\mathbf{v}$ and spatial position $\mathbf{x}$, and only consider reconstruction in velocity space, keeping $\mathbf{x}$ fixed. To simplify notation, we therefore
omit $\mathbf{x}$ and $t$. We also assume scaled variables, so that the molecular mass and Boltzmann constant do not appear in the subsequent equations.

We assume the following ansatz for the velocity distribution function (VDF) $f$:
\begin{equation}
    f(\mathbf{v}) \approx \sum_{i,j,k}f_{ijk} \delta\left(\mathbf{v} - \mathbf{v}_{ijk}\right).\label{eq:f-dvm}
\end{equation}
Here $\mathbf{v}_{ijk}$ are fixed velocity nodes, and $f_{ijk}$ are the unknown values of the distribution function. The nodes have associated quadrature weights $\Delta v_{ijk}$ that we use
when integrating functionals of $f$.
In the present work, we consider tensor-product uniform grids, although non-uniform grids can also be used, for example, those obtained using Wheeler's algorithm.
For a tensor-product uniform grid we have
\begin{equation}
    \mathbf{v}_{ijk} = \left(v_{x,\min} + (i-1) \Delta v_x, v_{y,\min} + (j-1) \Delta v_y, v_{z,\min} + (k-1) \Delta v_z  \right),
\end{equation}
\begin{equation}
    \Delta v := \Delta v_{ijk} = \Delta v_x \Delta v_y \Delta v_z.
\end{equation}
Here $v_{d,\min},\:d=x,y,z$ is the lower grid extent in the corresponding velocity direction, and $\Delta v_d,\: d=x,y,z$ is the grid spacing. The indices $i,j,k$ range from $1$ to $N_{v,x}$, $N_{v,y}$,
$N_{v,z}$, respectively. Thus, representation~(\ref{eq:f-dvm}) contains $N_{v,x}\times N_{v,y} \times N_{v,z}$ unknowns.
The expression for the moment $M_{k_x,k_y,k_z}$ becomes
\begin{equation}
    M_{k_x,k_y,k_z} = \Delta v \sum_i \sum_j \sum_k v_{x,i}^{k_x} v_{y,j}^{k_y} v_{z,k}^{k_z} f_{ijk},
\end{equation}
where $v_{x,i} = v_{x,\min} + (i-1) \Delta v_x$, $v_{y,j} = v_{y,\min} + (j-1) \Delta v_y$, $v_{z,k} = v_{z,\min} + (k-1) \Delta v_z$.

This can be succinctly re-written in a scalar product form by using an index mapping $\mathcal{L}$ from the set of triple indices to a single index given by
\begin{equation}
    l := \mathcal{L}(i,j,k) = i + N_{v,x}(j-1) + N_{v,x}N_{v,y}(k-1).
\end{equation}
The inverse mapping is then give by
\begin{equation}
    \footnotesize
    (i,j,k) = \mathcal{L}^{-1}(l) = \left(\left(l - 1 - N_{v,x}N_{v,y}\hat{k} \right) \bmod N_{v,x}+ 1,
    \left\lfloor \tfrac{l - 1 - N_{v,x}N_{v,y}\hat{k}}{N_{v,x}} \right\rfloor + 1,
    \hat{k} + 1\right),
\end{equation}
where $\hat{k} =  \left\lfloor \frac{l-1}{N_{v,x}N_{v,y}} \right\rfloor$, and $\lfloor \cdot \rfloor$ denotes the operation of rounding down to the nearest integer.
Using this single index $l$, we can ``unroll'' the 3-dimensional tensor $\mathbf{f}$ (with elements $f_{ijk})$ into a long vector $\mathbf{f}$ (with elements $f_l$), and defining a ``moment measurement vector''
\begin{equation}
    \mathbf{A}_{k_x,k_y,k_z} \in \mathbb{R}^{N_{v,x}\times N_{v,y} \times N_{v,z}} := \Delta v \sum_l v_{x,l}^{k_x} v_{y,l}^{k_y} v_{z,l}^{k_z}
\end{equation}
we then have the expression for a mixed-order moment as
\begin{equation}
    M_{k_x,k_y,k_z} = \left \langle \mathbf{A}_{k_x,k_y,k_z}, \mathbf{f} \right \rangle,
\end{equation}
where the angular brackets denote the standard scalar product.
This can be extended to multiple moments by introducing a ``moment measurement matrix'' $A_m$ that allows for computation of $m$ moments simultaneously via a single matrix-vector product:
\begin{equation}
    \mathbf{A}_m \in \mathbb{R}^{m \times \left(N_{v,x}\times N_{v,y} \times N_{v,z}\right)}:= \begin{pmatrix}
       \rule[.5ex]{2.5em}{0.4pt} & \mathbf{A}_{k_x^{(1)},k_y^{(1)},k_z^{(1)}}  & \rule[.5ex]{2.5em}{0.4pt} \\
       & \ldots & \\
       \rule[.5ex]{2.5em}{0.4pt} & \mathbf{A}_{k_x^{(m)},k_y^{(m)},k_z^{(m)}}  & \rule[.5ex]{2.5em}{0.4pt}
    \end{pmatrix}.\label{eq:measurement-matrix}
\end{equation}
This matrix then defines the set of moments given by the set of $m$ multi-indices $(k_x^{(1)},k_y^{(1)},k_z^{(1)}), \ldots,  (k_x^{(m)},k_y^{(m)},k_z^{(m)})$.

Therefore, given a set of $m$ moment values $M_{k_x^{(1)},k_y^{(1)},k_z^{(1)}}, \ldots, M_{k_x^{(m)},k_y^{(m)},k_z^{(m)}}$ and the associated moment measurement matrix $A_m$, which is dependent
on the quadrature nodes and weights used to for representation (\ref{eq:f-dvm}), we aim to find the values $f_l$ such that the following holds:
\begin{equation}
\mathbf{A}_m \mathbf{f} = \mathbf{M} := \begin{pmatrix}
        M_{k_x^{(1)},k_y^{(1)},k_z^{(1)}} \\
        \vdots \\
        M_{k_x^{(m)},k_y^{(m)},k_z^{(m)}}.
    \end{pmatrix}
\end{equation}

Having formalized the problem setting, we next discuss specific methods for finding the values $f$ that are either sparse or have a low-rank structure.

\subsection{Sparse entropic quadrature}\label{sec:speq}
The sparse entropic quadrature (which we will refer to as ``SPEQ'') is an extension of the entropic quadrature method (``EQMOM'') of Böhmer and Torrilhon~\cite{bohmer2020entropic}.
To recap, the entropic quadrature method proposes using a discretized representation of $f$ in the form~(\ref{eq:f-dvm}), and solving the constrained entropy minimization problem
\begin{equation}
  \begin{aligned}
      \min_{\substack{f_{l}}} \quad & \sum_{l} \Delta v_{l} f_{l} \log f_{l}\\
  \textrm{s.t.}  \quad & \mathbf{A}_m \mathbf{f} = \mathbf{M},\\
  \textrm{s.t.} \quad & f_{l} \ge 0.
  \end{aligned}\label{eq:optimization-eqmom}
\end{equation}
The constraint $f_{l} \ge 0$ comes from obvious physical considerations, as well as due to the logarithm function appearing in the objective. EQMOM avoids some of the instabilities
encountered by the classical maximum entropy method, which assumes a smooth $f$ and uses quadrature only for the evaluation of integrals, especially if the quadrature nodes and weights are chosen via Wheeler's
algorithm, although uniform grids also perform well.
The constrained optimization problem~(\ref{eq:optimization-eqmom}) usually has significantly less constraints than unknown variables, as one typically is interested in the lower-order moments of the distribution
function (on the order of 10-20), whereas the representation~(\ref{eq:f-dvm}) on a 3-dimensional 10$^3$ grid would already have 1000 values; finer grids are oftentimes used in kinetic solvers to capture
small-scale details crucial to the physics.
A standard approach for dealing with under-constrained optimization problems is their regularization via introduction of an $L_2$ or $L_1$ norm into the objective function. Since we are interested in obtaining
a sparse representation of $f$, $L_1$ would seem to be the obvious choice — however, due to the constraint $f_l \ge 0$ we have
\begin{equation}
    |\mathbf{f}|_1 = \sum_l |f_{l}| = \sum_l f_{l} = \rho.
\end{equation}
That is, the $L_1$ norm of the vector of unknowns $\mathbf{f}$ is simply the 0-order moment, i.e.\ density, which is obviously a conserved quantity that cannot be minimized.

We therefore propose using the following representation for $f$:
\begin{equation}
    f(\mathbf{v}) \approx \sum_{i,j,k}f_{ijk} \delta\left(\mathbf{v} - \mathbf{v}_{ijk}\right) = \sum_{i,j,k}w_{ijk} g_{ijk} \delta\left(\mathbf{v} - \mathbf{v}_{ijk}\right),\label{eq:f-dvm-weighted}
\end{equation}
where $w_{ijk}$ is a weighting function known at the start of the optimization procedure. We then can write the regularized optimization problem as
\begin{equation}
  \begin{aligned}
      \min_{\substack{g_{l}}} \quad & \sum_{l} \Delta v_{l} w_{l} g_{l} \log \left(w_{l} g_{l}\right) + \lambda |\mathbf{g}|_1\\
  \textrm{s.t.}  \quad & \hat{\mathbf{A}}_m \mathbf{g} = \mathbf{M},\\
  \textrm{s.t.} \quad & g_{l} \ge 0.
  \end{aligned}\label{eq:optimization-spec}
\end{equation}
This corresponds to the constrained entropy minimization with additional \textit{weighted} $L_1$ regularization~\cite{candes2008enhancing}, whose strength is governed by the value of $\lambda$. With $\lambda=0$, the original EQMOM is recovered, albeit
in a setting where one solves for $g_{ijk}$ (if $w_{ijk} \equiv 1$, then the EQMOM setting is fully recovered). Use of the weighting $w_{ijk}$ is also beneficial for numerical stability — typical kinetic distributions vary over many orders of magnitude, which might lead to numerical issues and slow convergence if one tries to to find these values directly. By using an appropriate weighting,
the range of values of $g_{ijk}$ can be significantly narrower, thus improving the performance of the algorithm.
Here $\hat{\mathbf{A}}_m$ is defined similarly to~(\ref{eq:measurement-matrix}), as a concatenation of measurement
vectors $\hat{\mathbf{A}}_{k_x,k_y,k_z}$ that incorporate the weighting:
\begin{equation}
\hat{\mathbf{A}}_{k_x,k_y,k_z} := \Delta v \sum_l w_l v_{x,l}^{k_x} v_{y,l}^{k_y} v_{z,l}^{k_z}.\label{eq:measurement-matrix-weighted}
\end{equation}
We then can compute a mixed-order moment as a function of $\mathbf{g}$ as
\begin{equation}
    M_{k_x,k_y,k_z} = \left \langle \hat{\mathbf{A}}_{k_x,k_y,k_z}, \mathbf{g} \right \rangle.
\end{equation}

In case no prior information on $f_{ijk}$ is available, and one knows only the moments, then an obvious choice of weighting is the local Maxwellian distribution, given by
\begin{equation}
    w_{ijk}(T) = M_{0,0,0} \frac{1}{n_{MB}(T)}\exp\left(-\frac{||\mathbf{v}_{ijk}||_2^2}{T} \right),
\end{equation}
where the density $n_{MB}$ is given by
\begin{equation}
    n_{MB}(T) = \sum_{ijk} \Delta v_{ijk} \exp\left(-\frac{||\mathbf{v}_{ijk}||_2^2}{T} \right).
\end{equation}
It should be noted that here we assume that the mean velocity is 0. In case it is not, find the local Maxwellian assuming
no streaming velocity, and then shift the resulting
distribution by the actual mean velocity as given by the first-order moments $M_{1,0,0}$, $M_{0,1,0}$, $M_{0,0,1}$, thus obtaining the weighting $w_{ijk}$.
The number density and temperature $T$ of the Maxwellian distribution are found from the lowest-order moments. It should be noted that due to the discretization~(\ref{eq:f-dvm}), the standard gas-kinetic relationship $e=\frac{3}{2}T$ does not hold, as the discretization introduces an error. Therefore, the procedure to finding $w_{ijk}$ is the following:
\begin{enumerate}
    \item Given the moments $M_{k_x^{(1)},k_y^{(1)},k_z^{(1)}}, \ldots, M_{k_x^{(m)},k_y^{(m)},k_z^{(m)}}$, compute the density of the Maxwellian as $M_{0,0,0}$ and the grid offset velocity $\overline{\mathbf{v}}$ as \newline $\frac{1}{M_{0,0,0}}(M_{1,0,0}, M_{0,1,0}, M_{0,0,1})$
    \item Compute the energy as $e = (\frac{M_{2,0,0}}{M_{0,0,0}} - \overline{v}_{x})^2 + (\frac{M_{0,2,0}}{M_{0,0,0}} - \overline{v}_{y})^2 + (\frac{M_{0,0,2}}{M_{0,0,0}} - \overline{v}_{z})^2$
    \item Estimate $T_0$ as $\frac{2}{3}e$
    \item Find $T$ by solving the non-linear equation using Newton's method with an initial guess of $T_0$:
    \begin{equation}
        \frac{1}{\sum_{ijk} \Delta v_{ijk} w_{ijk}(T)} \sum_{ijk} \Delta v_{ijk} ||\mathbf{v}_{ijk}||_2^2 w_{ijk}(T) = e
    \end{equation}
    \item Compute $w_{ijk}(\overline{v}_{x}, \overline{v}_{y}, \overline{v}_{z}, T)$ as a shifted Maxwellian
\end{enumerate}

Having obtained the weighting function $w_{ijk}(\overline{v}_{x}, \overline{v}_{y}, \overline{v}_{z}, T)$, we proceed to solve~(\ref{eq:optimization-spec}). But first we generalize the objective appearing in~(\ref{eq:optimization-spec}), replacing the discretized entropy
with the Kullback--Leibler (KL) divergence~\cite{kullback1951information} with respect to a known distribution $\phi_{ijk}$ (with the
corresponding unrolled vector of values $\phi_l$):
\begin{equation}
  \begin{aligned}
      \min_{\substack{g_{l}}} \quad & \sum_{l} \Delta v_{l} w_{l} g_{l} \log \left(\frac{w_{l} g_{l}}{\phi_l}\right) + \lambda |\mathbf{g}|_1\\
  \textrm{s.t.}  \quad & \hat{\mathbf{A}}_m \mathbf{g} = \mathbf{M},\\
  \textrm{s.t.} \quad & g_{l} \ge 0.
  \end{aligned}\label{eq:optimization-spec-KL}
\end{equation}
In case $\phi_{ijk}$ is not known, we can take it as $\phi_{ijk} \equiv 1$ and recover the sparse entropic quadrature method~(\ref{eq:optimization-spec}). However, in case we are trying to store a sparse representation of a known
distribution, for example, obtained via a high-resolution kinetic solver, we can use it in~(\ref{eq:optimization-spec-KL}) to minimize the discrepancy between the sparse representation $f_{ijk}$ and the original
distribution $\phi_{ijk}$ by optimizing the Kullback--Leibler divergence, a measure of relative entropy. A discussion of other possible divergences in the framework of kinetic theory can be found in~\cite{abdelmalik2016moment}.
 It can be seen from the optimization objective in~(\ref{eq:optimization-spec-KL}) that
$g_l$ cannot be exactly equal to 0 due to the logarithm, therefore, for practical purposes, we set negligibly small values of $g$ to 0 --- that is, values smaller than a certain threshold $\varepsilon$ are taken to be 0. The impact of the choice
of $\varepsilon$ is discussed in the numerical results section.

We now discuss the numerical solution of~(\ref{eq:optimization-spec-KL}). It can be conveniently formulated
as an unconstrained dual problem, reducing the number of variables significantly, as it is reasonable to assume that the number of moment constraints $M$ is significantly less than $N_{v,x}\times N_{v,y} \times N_{v,z}$, the latter being the number of values of $g$ we aim to find.
The Lagrange function of the problem is given by
\begin{equation}
    \mathcal{L}(\mathbf{y}, \mathbf{g}) = \sum_{l} \Delta v_{l} w_{l} g_{l} \log \left(\frac{w_{l} g_{l}}{\phi_l}\right) + \lambda |\mathbf{g}|_1 + \mathbf{y}^T \left(\hat{\mathbf{A}}_m \mathbf{g} - \mathbf{M}_m\right),\label{eq:lagrangian}
\end{equation}
where $\mathbf{y} \in \mathbb{R}^m$.

Let us denote the i-th component of the vector $\hat{\mathbf{A}}_m^T \mathbf{y}$ as $s_i$.
From the optimality condition $\partial \mathcal{L}/\partial g_i = 0$ we obtain
\begin{equation}
    \Delta v_{i} w_{i} \left(\log \left(\frac{w_{i} g_{i}}{\phi_i}\right) + 1\right) + \lambda + s_i = 0,
\end{equation}
from which we find $g_i$ in terms of $\mathbf{y}$ (since $s_i = s_i(\mathbf{y})$):
\begin{equation}
    g_i = \frac{\phi_i}{w_i} z_i(\mathbf{y}),\label{eq:g_i_from_y}
\end{equation}
where \begin{equation}
    z_i(\mathbf{y}) = \exp\left(-1 - \frac{\lambda + s_i}{\Delta v_{i} w_{i}}\right).
\end{equation}

Plugging the solution into~(\ref{eq:lagrangian}), one obtains the dual unconstrained optimization problem:
\begin{equation}
    \max \left( -\mathbf{y}^T \mathbf{M}_m  - \sum_l \Delta v_l \phi_l z_l \right) =: \max \Phi(\mathbf{y}).\label{eq:dual-targ}
\end{equation}
To solve the unconstrained dual problem, a Newton method with Armijo backtracking line search is used, with the
gradient given by
\begin{equation}
    \nabla \Phi = - \mathbf{M}_m + \hat{\mathbf{A}}_m \mathbf{d},\label{eq:dual-grad}
\end{equation}
where $d_l =  \phi_l  z_l / w_l$.
The gradient $\nabla \Phi$ depends on $\mathbf{y}$ only through the terms $\mathbf{z}(\mathbf{y})$ appear in $\mathbf{d}$. The Hessian of $\Phi$ can be computed to be equal to
\begin{equation}
    \nabla^2 \Phi = \hat{\mathbf{A}}_m \mathbf{H} \hat{\mathbf{A}}_m^T,\label{eq:dual-hess}
\end{equation}
where $\mathbf{H} \in \mathbb{R}^{\left(N_{v,x}\times N_{v,y} \times N_{v,z}\right) \times \left(N_{v,x}\times N_{v,y} \times N_{v,z}\right)}$ is a diagonal matrix with diagonal entries defined via
\begin{equation}
    H_{ll} = -\frac{\phi_l z_l}{\Delta v_{l} w_l^2}.
\end{equation}

The Newton method is initialized with a solution $\mathbf{y}_0$ obtained from~(\ref{eq:g_i_from_y}) with $g_i \equiv 1$, and assuming $\lambda = 0$. This corresponds to starting the non-weighted entropic quadrature method with
the weighting distribution function $w$.
From~(\ref{eq:g_i_from_y}) with $g_i \equiv 1$ and $\lambda = 0$ we obtain the following system of linear equations to solve to compute the initial solution $\mathbf{y}_0$:
\begin{equation}
    \hat{\mathbf{A}}_{m} \hat{\mathbf{A}}^T_{m} \mathbf{y}_0 = -\hat{\mathbf{A}}_{m} \mathbf{r},\label{eq:dual-init}
\end{equation}
where
\begin{equation}
    r_l = w_l \Delta v_l  \left(\log\left( \frac{\phi_l}{w_l}  \right) -1 \right).
\end{equation}

Finally, since the condition number of $\hat{\mathbf{A}}_m$ directly affects the condition number of the Hessian $\mathbf{H}$ of the dual problem,
we scale $\hat{\mathbf{A}}_m$. We define
\begin{equation}
    \mathbf{D}_r \in \mathbb{R}^{m \times m} := \mathrm{diag}\left(||\hat{\mathbf{A}}_{m,1}||_2,\ldots,||\hat{\mathbf{A}}_{m,m}||_2 \right),
\end{equation}
where $||\hat{\mathbf{A}}_{m,i}||_2$ denotes the 2-norm of row $i$ of $\hat{\mathbf{A}}_m$.
Left-multiplying the constraint $\hat{\mathbf{A}}_m \mathbf{g} = \mathbf{M}_m$ by the inverse of $\mathbf{D}_r$,
we can write it as
\begin{equation}
    \hat{\mathbf{A}}_{s} \mathbf{g} = \mathbf{M}_s,
\end{equation}
where $\hat{\mathbf{A}}_{s}$ = $\mathbf{D}_r^{-1} \hat{\mathbf{A}}_m$, $\mathbf{M}_s = \mathbf{D}_r^{-1} \mathbf{M}$.
The scaled matrix $\hat{\mathbf{A}}_{s}$ and vector $\mathbf{M}_s$ are then used instead of
$\hat{\mathbf{A}}_{m}$ and $\mathbf{M}_m$ in the definition of $s_i$ and Eqns.~(\ref{eq:dual-targ}), (\ref{eq:dual-grad}), (\ref{eq:dual-hess}), and (\ref{eq:dual-init}).
This concludes the section on the sparse entropic quadrature method.

\begin{algorithm}
\caption{Sparse entropic quadrature for moment closure}
\label{alg:sparse_entropic_momentclosure}

 \hspace*{\algorithmicindent} \textbf{Input:} Moments $\mathbf{M}_m$ of order up to $m$, regularization $\lambda$, threshold $\varepsilon$, velocity nodes $\mathbf{v}_{ijk}$, quadrature weights $\Delta v_{ijk}$ \\
 \hspace*{\algorithmicindent} \textbf{Output:} Next-order moments

\begin{algorithmic}[1]


\Statex \textbf{Phase 1: Weighting Function Initialization}
\State Compute mean velocity $\overline{\mathbf{v}}$ and energy $e$ from low-order moments
\State Solve for temperature $T$ using Newton's method:
    \Statex \hskip\algorithmicindent $\frac{1}{\sum w_{ijk}(T)} \sum ||\mathbf{v}_{ijk}||_2^2 w_{ijk}(T) = e$
\State Compute weighting function $w_{ijk}$ as a shifted Maxwellian using $\overline{\mathbf{v}}$ and $T$

\Statex \textbf{Phase 2: Dual Problem Initialization}
\State Compute the measurement matrix $\hat{\mathbf{A}}_m$ according to~(\ref{eq:measurement-matrix-weighted})
\State Compute $r_l = w_l \Delta v_l  \left(\log\left( \frac{\phi_l}{w_l}  \right) -1 \right)$
\State Initialize dual variables $\mathbf{y}_0$ by solving the linear system~(\ref{eq:dual-init}):
    \Statex \hskip\algorithmicindent $\hat{\mathbf{A}}_{m} \hat{\mathbf{A}}^T_{m} \mathbf{y}_0 = -\hat{\mathbf{A}}_{m} \mathbf{r}$

\Statex \textbf{Phase 3: Dual Space Optimization}
\While{not converged}
    \State Compute vector $\mathbf{s} = \hat{\mathbf{A}}_m^T \mathbf{y}$
    \State Compute $z_l(\mathbf{y}) = \exp\left(-1 - \frac{\lambda + s_l}{\Delta v_{l} w_{l}}\right)$
    \State Compute dual objective $\Phi(\mathbf{y}) = -\mathbf{y}^T \mathbf{M}_m  - \sum \Delta v_l z_l(\mathbf{y})$
    \State Update $\mathbf{y}$ using \textbf{Newton's Method}
    \State Perform \textbf{Armijo backtracking line search} for step size stability
\EndWhile

\Statex \textbf{Phase 4: Primal Recovery and Sparsification}
\State Compute $g_l$ from $\mathbf{y}$:
    \Statex \hskip\algorithmicindent $g_l = \frac{1}{w_l} \exp\left(-1 - \frac{\lambda + s_l}{\Delta v_{l} w_{l}}\right)$
\State \textbf{Thresholding:} Set $g_l = 0$ for all $g_l < \varepsilon$
\State Compute measurement matrix $\hat{\mathbf{A}}_{m_{\mathrm{next}}}$ for next-order moments
\State \Return $\hat{\mathbf{A}}_{m_{\mathrm{next}}} \mathbf{g}$

\end{algorithmic}
\end{algorithm}

\begin{algorithm}
\caption{Sparse entropic quadrature for sparse VDF reconstruction}
\label{alg:sparse_entropic_reconstruct}

 \hspace*{\algorithmicindent} \textbf{Input:} Known distribution function $f$, maximum order $m$ of moments to preserve, regularization $\lambda$, threshold $\varepsilon$, velocity nodes $\mathbf{v}_{ijk}$, quadrature weights $\Delta v_{ijk}$ \\
 \hspace*{\algorithmicindent} \textbf{Output:} Sparse representation of VDF $f$

\begin{algorithmic}[1]


\Statex \textbf{Phase 1: Initialization}
\State Define weighting function $w_{ijk}$ as $f_{ijk}$
\State Define target distribution $\phi_{ijk}$ as $f_{ijk}$

\Statex \textbf{Phase 2: Dual Problem Initialization}
\State Compute the measurement matrix $\hat{\mathbf{A}}_m$ according to~(\ref{eq:measurement-matrix-weighted})
\State Compute moments $\mathbf{M}_m$ of order up to $m$
\State Compute $r_l = w_l \Delta v_l  \left(\log\left( \frac{\phi_l}{w_l}  \right) -1 \right)$
\State Initialize dual variables $\mathbf{y}_0$ by solving the linear system~(\ref{eq:dual-init}):
    \Statex \hskip\algorithmicindent $\hat{\mathbf{A}}_{m} \hat{\mathbf{A}}^T_{m} \mathbf{y}_0 = -\hat{\mathbf{A}}_{m} \mathbf{r}$

\Statex \textbf{Phase 3: Dual Space Optimization}
\While{not converged}
    \State Compute vector $\mathbf{s} = \hat{\mathbf{A}}_m^T \mathbf{y}$
    \State Compute $z_l(\mathbf{y}) = \exp\left(-1 - \frac{\lambda + s_l}{\Delta v_{l} w_{l}}\right)$
    \State Compute dual objective $\Phi(\mathbf{y}) = -\mathbf{y}^T \mathbf{M}_m  - \sum \Delta  v_l \phi_l z_l(\mathbf{y})$
    \State Update $\mathbf{y}$ using \textbf{Newton's Method}
    \State Perform \textbf{Armijo backtracking line search} for step size stability
\EndWhile

\Statex \textbf{Phase 4: Primal Recovery and Sparsification}
\State Compute $g_l$ from $\mathbf{y}$:
    \Statex \hskip\algorithmicindent $g_l = \frac{\phi_l}{w_l} \exp\left(-1 - \frac{\lambda + s_l}{\Delta v_{l} w_{l}}\right)$
\State \textbf{Thresholding:} Set $g_l = 0$ for all $g_l < \varepsilon$
\State \Return $f(\mathbf{v}) \approx \sum w_l g_l \delta(\mathbf{v} - \mathbf{v}_l)$

\end{algorithmic}
\end{algorithm}

The two versions of the SPEQ method, i.e.\ the entropy minimization-based solution of the moment closure problem and the Kullback--Leibler divergence minimization-based
sparse reconstruction of the distribution function, are summarized in Algorithms~\ref{alg:sparse_entropic_momentclosure} and~\ref{alg:sparse_entropic_reconstruct}, respectively.

\subsection{Low-rank constrained moment recovery}\label{sec:mclr}

The low-rank constrained moment recovery (LRMR) approach is based on the observation that an equilibrium Maxwell--Boltzmann distribution is separable in the three velocity directions and therefore has Tucker rank $(1,1,1)$. Distributions close to equilibrium, or distributions with a small number of dominant kinetic structures, can thus be represented by a low-rank tensor while preserving prescribed moments.
We use the Tucker ansatz (see, e.g.,~\cite{koldaTensorDecompositionsApplications2009}),
\begin{equation}
    f_{ijk} \approx \sum_{\alpha=1}^{r_x}\sum_{\beta=1}^{r_y}\sum_{\gamma=1}^{r_z}
    C_{\alpha\beta\gamma} U^{(x)}_{i\alpha} U^{(y)}_{j\beta} U^{(z)}_{k\gamma},
\end{equation}
and in the numerical experiments take $r_x=r_y=r_z=r$. Let $\mathcal{T}_r$ denote the set of tensors whose mode-$x$, mode-$y$, and mode-$z$ unfoldings have rank at most $r$. We then solve the constrained least-squares problem
\begin{equation}
  \begin{aligned}
      \min_{\mathbf{f} \in \mathcal{T}_r} \quad & \frac{1}{2}\left\|\mathbf{A}_{s}\mathbf{f}-\mathbf{M}_{s}\right\|_2^2\\
      \textrm{s.t.} \quad & f_{ijk} \geq 0,
  \end{aligned}\label{eq:optimization-mclr}
\end{equation}
where $\mathbf{A}_{s}=\mathbf{D}_r^{-1}\mathbf{A}_m$ and $\mathbf{M}_{s}=\mathbf{D}_r^{-1}\mathbf{M}$ are obtained by the same row scaling as in the SPEQ method, see Section~\ref{sec:speq}. The scaling is important because the rows corresponding to high-order moments can differ by several orders of magnitude from the rows corresponding to low-order moments.

Problem~(\ref{eq:optimization-mclr}) is solved by projected gradient descent. Given $\mathbf{F}^n$, we compute
$\mathbf{r}^n=\mathbf{A}_s\operatorname{vec}(\mathbf{F}^n)-\mathbf{M}_s$ and
$\mathbf{G}^n=\operatorname{reshape}(\mathbf{A}_s^T\mathbf{r}^n)$, and set
\begin{equation}
    \mathbf{F}^{n+1} =
    \max\left\{\mathcal{P}_r\left(\mathbf{F}^n-\tau_n \mathbf{G}^n\right),0\right\}.
\end{equation}
Here $\mathcal{P}_r$ denotes the Tucker-rank-$r$ projection, computed by a sequentially truncated higher-order SVD. The step size $\tau_n$ is selected by backtracking: if the trial iterate increases $\|\mathbf{A}_s\operatorname{vec}(\mathbf{F})-\mathbf{M}_s\|_2$, the step is reduced and the trial is recomputed. If no descent step is found after $N_{\textnormal{BT}}$ of backtracking steps, the iteration is terminated. For the numerical results presented in Section~\ref{sec:numres}, we take $N_{\textnormal{BT}}=30$ and an initial step size equal to the inverse of the spectral norm of $\mathbf{A}_s$.

The initialization can be chosen from an available reference distribution or, when only moments are known, from the Maxwellian constructed from the density and temperature as described above. In the implementation, the initial tensor is first projected to $\mathcal{T}_r$ and then clipped to enforce nonnegativity. Since the rank and nonnegativity projections are applied successively, we can not expect to find a global minimizer of~(\ref{eq:optimization-mclr}) and there are multiple local minima. The formulation nevertheless directly penalizes moment mismatch while keeping the number of stored degrees of freedom small.

For a given rank $r$ and assuming the same number of grid nodes $n_v$ per direction, the total number of degrees of freedom is equal to $r^3 + 3r n_v$.

\section{Numerical results}\label{sec:numres}
To test the developed approaches, we apply them to the non-dynamical problem of reconstructing an unknown 3-dimensional velocity distribution function
based on a limited number of moment measurements with the moments computed using a ``hidden truth'' distribution. This allows us to
compute errors with respect to the ``hidden truth'' distribution.
For all cases, we consider moment constraints (i.e.\ moments that are to be conserved) of the form
$M_{k_x,k_y,k_z}: k_x+k_y+k_z \leq m_{\mathrm{max}}$, and choose $m_{\mathrm{max}}=4,6$.

As a measure of error, we consider the
\begin{itemize}
    \item relative $L_2$ error in the reconstructed velocity distribution function with respect to the hidden truth distribution
    \item the absolute and relative errors $E_{m_{\mathrm{max}}+1}^{\mathrm{abs}}$, $E_{m_{\mathrm{max}}+1^{\mathrm{rel}}}$ in the next predicted moments compared with those given by the hidden truth distribution.
\end{itemize}
We define the absolute error $E_{m_{\mathrm{max}}+1}^{\mathrm{abs}}$ as
\begin{equation}
    E_{m_{\mathrm{max}}+1}^{\mathrm{abs}} = \sum_{k_x+k_y+k_z=m_{\mathrm{max}}+1} \left|M_{k_x,k_y,k_z}^{\mathrm{pred}} - M_{k_x,k_y,k_z}^{\mathrm{true}}\right|,
\end{equation}
and the relative error $E_{m_{\mathrm{max}}+1}^{\mathrm{rel}}$ as
\begin{equation}
    E_{m_{\mathrm{max}}+1}^{\mathrm{rel}}
     = \frac{\sum_{k_x+k_y+k_z=m_{\mathrm{max}}+1} \left|M_{k_x,k_y,k_z}^{\mathrm{pred}}
      - M_{k_x,k_y,k_z}^{\mathrm{true}}\right|}{\sum_{k_x+k_y+k_z=m_{\mathrm{max}}+1} \left|M_{k_x,k_y,k_z}^{\mathrm{true}}\right|}.
\end{equation}
Here $M_{k_x,k_y,k_z}^{\mathrm{pred}}$ are the values of the moments computed using the reconstructed VDF, whereas
$M_{k_x,k_y,k_z}^{\mathrm{pred}}$ are computed using the hidden truth distribution. The reason for the use of both the relative and absolute errors is that for some of the distributions considered,
specifically the Maxwell-Boltzmann and Druyvesteyn distributions,
the values of the moments of the hidden truth distribution are zero, and the relative error cannot be used.

The data used in the present work has been made publicly available on Zenodo~\cite{otwth2026sparsedata} and includes the values of the hidden
truth distributions, as well as the predictions obtained via the different approaches developed in the present work; the data post-processing script used to produce the plots is also included.
The NuFI data for collisionless plasmas used as ground truth data in Sec.~\ref{sec:collisionlessplasmas} is available on Zenodo~\cite{wilhelm2026beamdata}.
The codes implementing the SPEQ approach has been made publicly available on GitHub~\cite{oblapenko2026specqk}; the simulation setups and post-processing scripts can be found in the \\\texttt{simulations/paper\_2026\_sparse\_and\_lowrank} directory of the
repository, along with details on the steps required to run the simulations in \texttt{README.md}. Results presented in the
current work were produced using version 0.1.0 of the code.
The low-rank constrained moment recovery code is publicly available~\cite{theisen2026sparsetensor}.

\subsection{Model distributions}
First, we consider several model distributions from gas dynamics problems.
For the SPEQ approach, we take the following $\lambda$ values:
\begin{equation}
    \tilde{\lambda} := \frac{\lambda}{\Delta v} \in \{0, 10^{-10},  10^{-8}, 10^{-6}, 10^{-5}, 10^{-4}, 10^{-3}, 10^{-2}, 0.1, 0.5\}
\end{equation}
where $\Delta v$ is the quadrature weight; since we consider uniform grids, the quadrature weights are constant across
all nodes.
Unless stated otherwise, in the SPEQ approach, values of $g$ that are smaller than $\varepsilon = 10^{-7}$ were set to 0; a parameter study of the impact of the choice of $\varepsilon$ on
the errors and solution sparsity is performed for the Mott--Smith distribution.

\subsubsection{Maxwell--Boltzmann distribution}
We start with the equilibrium Maxwell--Boltzmann distribution, computed as
\begin{equation}
    f(\mathbf{v}) = C_{MB} \exp\left(-\frac{||\mathbf{v}||_2^2}{T} \right),
\end{equation}
with the factor $C_{MB}$ chosen so that the total number density is equal to 1; the temperature $T$ is also taken to be 1.
The distribution is evaluated on a uniform 40$^3$ grid with extent from $-4$ to $4$ in each velocity direction.

\begin{figure}[t!]
  \centering
  \includegraphics[width=0.85\textwidth]{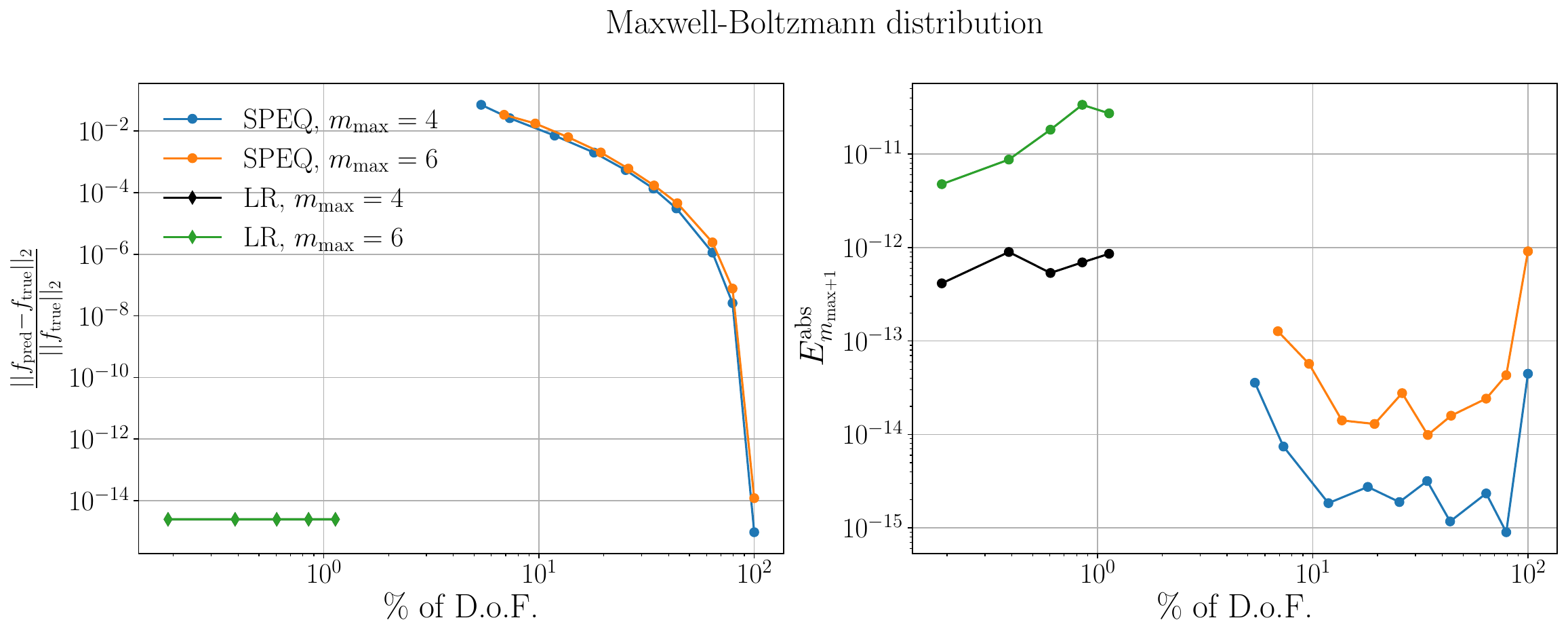}
  \caption{Relative $L_2$ error in the VDF (left) and absolute error in next predicted moments (right) as a function
  of the fraction of used degrees of freedom, Maxwell-Boltzmann distribution.}\label{fig:mb-errors}
  \end{figure}
\begin{figure}[t!]
  \centering
  \includegraphics[width=0.85\textwidth]{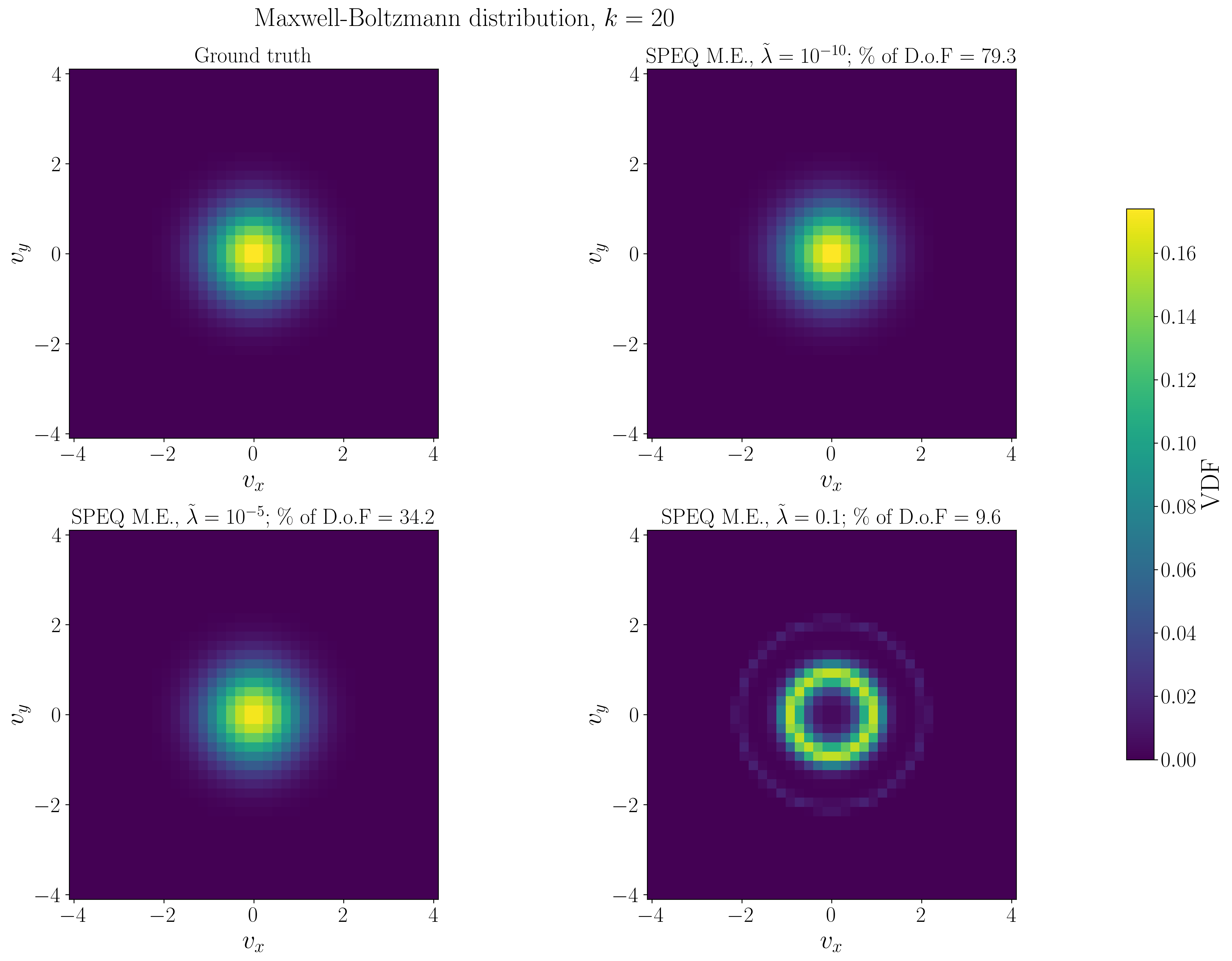}
  \caption{Slices along the $v_z$ axis ($k=20$) of the hidden truth distribution and distributions reconstructed using the SPEQ method for various values of $\lambda$, Maxwell-Boltzmann distribution. }\label{fig:mb-vzslice}
  \end{figure}
Figure~\ref{fig:mb-errors} shows the relative $L_2$ error (left) and error in next predicted moments (right) as a
function of the fraction of total number of degrees of freedom used to represent the VDF, with
the total number of degrees of freedom being $40^3 = 64000$. That is, in the SPEQ approach, higher regularization strengths $\lambda$ lead to stronger sparsity and fewer degrees of freedom used and thus
correspond to values on the left-hand side of the plots. In the LRMR approach, a lower rank $r$ leads to a lower number of degrees of freedom; in general, the number of degrees of freedom
required to store the distributions produced by LRMR is an order of magnitude lower than required for the SPEQ approach.
Nevertheless, very high levels of sparsity can be achieved within SPEQ, at the cost of introducing error in the distribution; however,
certain symmetry is retained, and therefore the next-order moments, which are odd, remain zero almost to machine precision. Since the Maxwell--Boltzmann distribution can be decomposed exactly into
a product of rank-1 tensors, the LRMR method exhibits virtually no error for any rank $r \geq 1$.

We plot a slice of the distribution along the $v_z$ axis in Figure~\ref{fig:mb-vzslice} for various values of $\lambda$, along with the slice of the underlying true distribution, when all moments up
to total order 6 are conserved.
For the slice, the index $k$ was taken as $20$, which on a symmetric $40^3$ grid corresponds to the smallest positive $v_z$ velocity value.
We observe very little visual difference between the reference distribution and the reconstructed values except for a very large value of $\lambda$, which corresponds to a very sparse solution, and leads
a ``hollowed-out'' appearance in the distribution. This is due to the fact that the moment constraints remain satisfied, which simultaneously leads to
sparsification of the high-velocity tails of the distribution and to a redistribution of density from the lower velocity bulk to somewhat higher velocities to compensate for the reduction in the values of the higher-order moments caused by the induced sparsity in the distribution tails.
For very sparse representations, this leads to the ``hollowing-out'' seen above.

\subsubsection{Druyvesteyn distribution}
Next, we consider the Druyvesteyn distribution that arises in collisional plasma physics and is computed as
\begin{equation}
    f(\mathbf{v}) = C_{D} \exp\left(-\alpha \frac{||\mathbf{v}||_2^4}{T} \right),
\end{equation}
where
\begin{equation}
    \alpha = \left(\frac{2}{3T}\frac{\Gamma \left(5/4\right)}{\Gamma \left(3/4\right)}\right)^2.
\end{equation}
Similarly to the previous case, the temperature $T$ is also taken to be 1,
the distribution is evaluated on a uniform 40$^3$ grid with extent from $-4$ to $4$ in each velocity direction, and
the factor $C_{D}$ is chosen so that the total number density is equal to 1.

\begin{figure}[t!]
  \centering
  \includegraphics[width=0.85\textwidth]{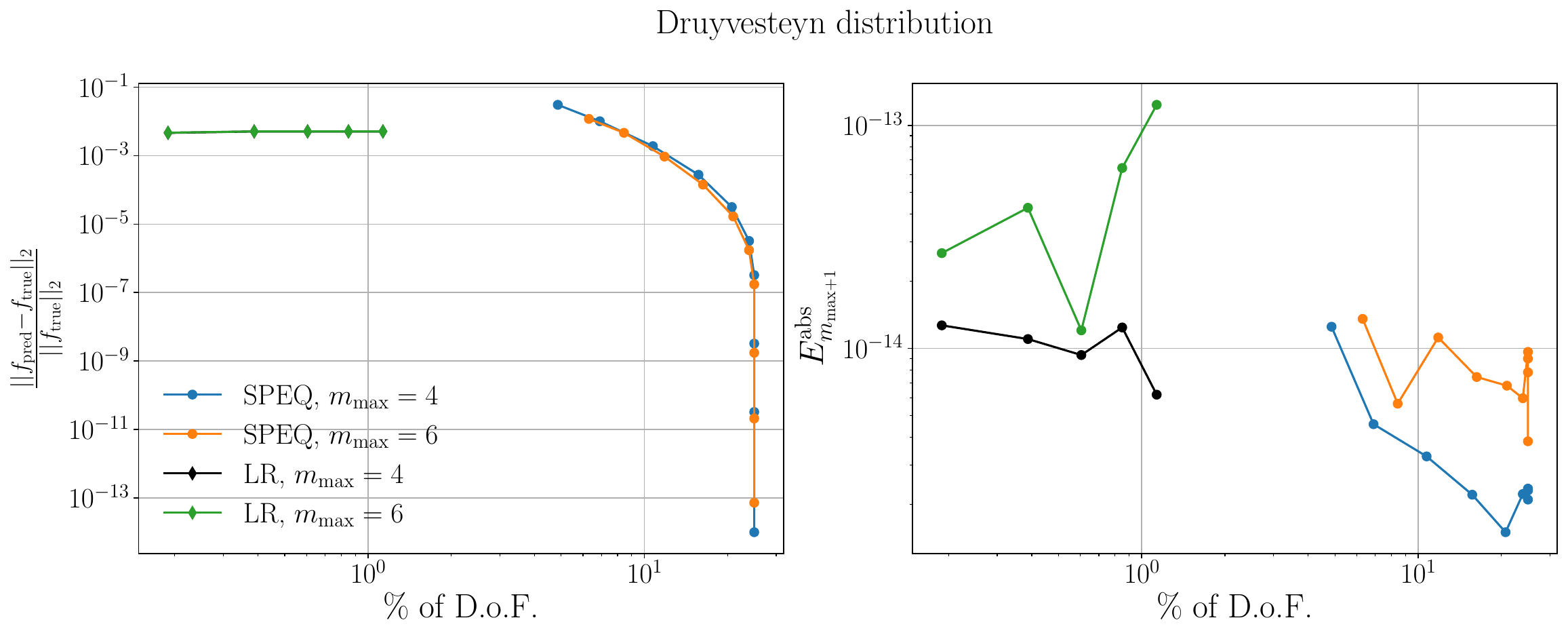}
  \caption{Relative $L_2$ error in the VDF (left) and absolute error in next predicted moments (right) as a function
  of the fraction of used degrees of freedom, Druyvesteyn distribution.}\label{fig:dr-errors}
  \end{figure}
  \begin{figure}[t!]
  \centering
  \includegraphics[width=0.85\textwidth]{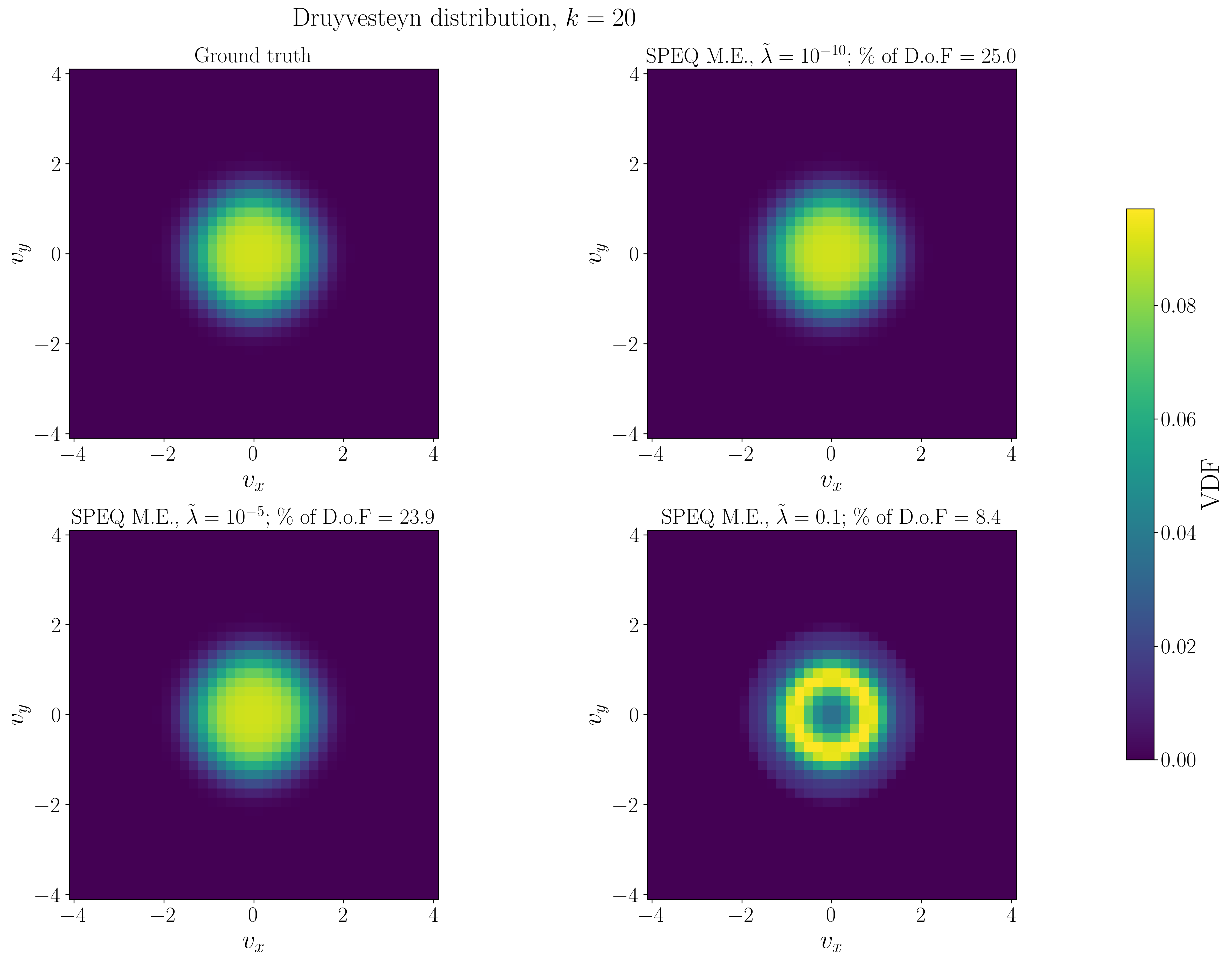}
  \caption{Slices along the $v_z$ axis ($k=20$) of the hidden truth distribution and distributions reconstructed using the SPEQ method for various values of $\lambda$, Druyvesteyn distribution. }\label{fig:dr-vzslice}
  \end{figure}
  \begin{figure}[h!]
  \centering
  \includegraphics[width=0.85\textwidth]{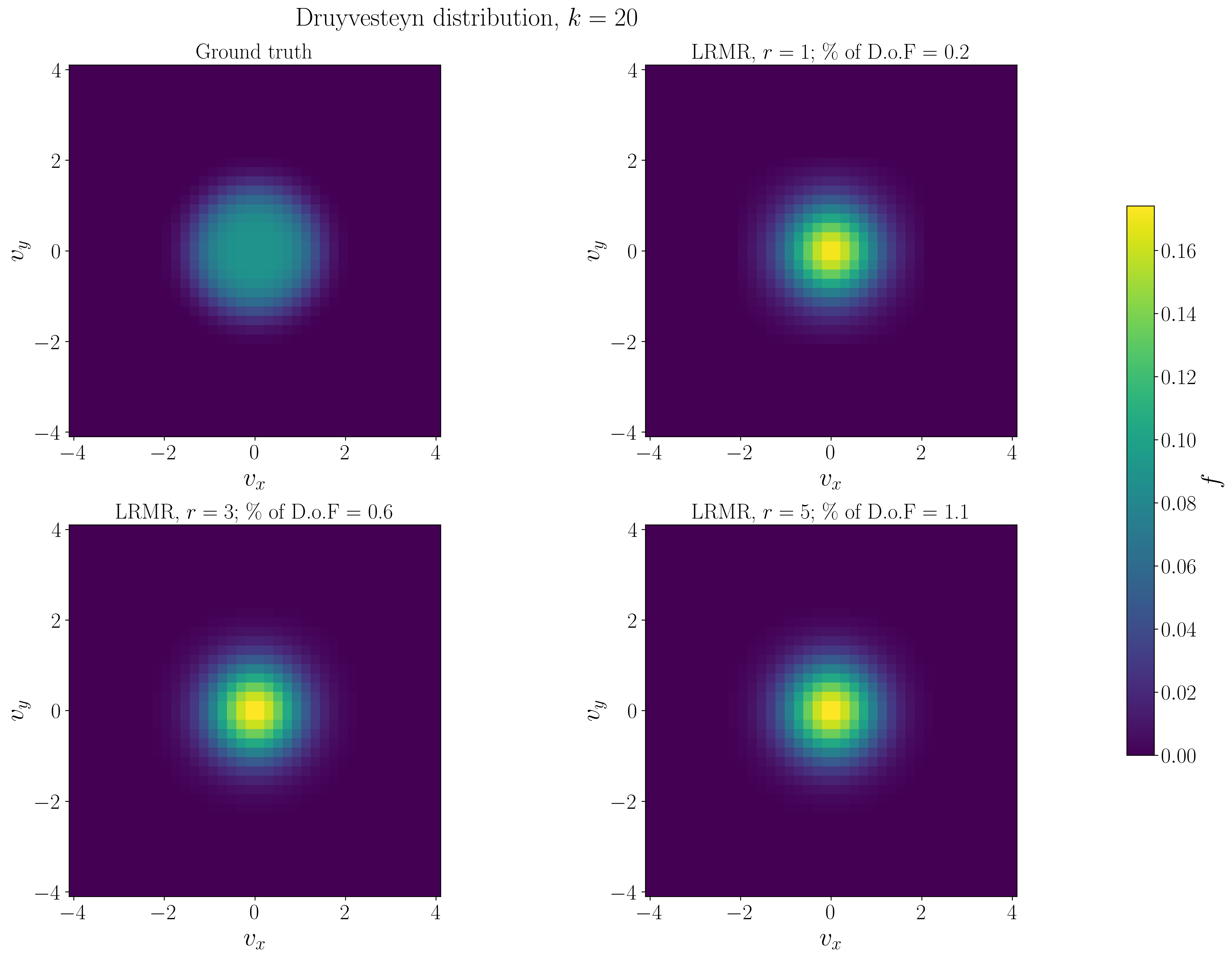}
  \caption{Slices along the $v_z$ axis ($k=20$) of the hidden truth distribution and distributions reconstructed using the LRMR method for various ranks $r$, Druyvesteyn distribution. }\label{fig:dr-lrmr-vzslice}
  \end{figure}
Figure~\ref{fig:dr-errors} shows the errors in the reconstructed distribution and the next predicted moments as a function of the fraction of used degrees of freedom. Due to the very fast decay of the distribution, even for $\lambda=0$, applying
a threshold of $10^{-7}$ to the values $g$ produced by the SPEQ approach leads to non-negligible sparsity in the solution,
and increasing $\lambda$ moderately has little effect on the number of degrees of freedom required to represent the VDF.
Due to the symmetry of the distribution, the analytical next-order moments are zero, and since the symmetry is well-retained regardless of
the values of $\lambda$, this is reflected in the very low absolute errors in the predicted values of the unknown moments, as seen on
the right-hand side of~\ref{fig:dr-errors}. The same holds for the LRMR approach, which retains the symmetry of the distribution; the relative $L_2$ error is however non-negligible, but
for the same level of error as that obtained via SPEQ, an order of magnitude fewer degrees of freedom are required.
We also plot a slice of the distribution predictions along the $v_z$ axis in Figure~\ref{fig:dr-vzslice} for $k=20$ and all moments up to total order 6 conserved.
Similarly to the Maxwell-Boltzmann distribution case, we notice very little visual difference between the reference distribution and the reconstructed values, unless $\lambda$ is taken quite large;
as in the previous case, a ``hollowed-out'' appearance in the distribution for high sparsity values.

Finally, Figure~\ref{fig:dr-lrmr-vzslice} shows a slice of the distributions obtained using the LRMR method when all moments up to order 6 are conserved. We see that it predicts larger values of the distribution,
but retains the overall symmetry.

\subsubsection{Mott--Smith distribution}
Finally, we consider a 3-dimensional version of the Mott--Smith distribution~\cite{mott1951solution} describing the velocity distribution before, inside, and after a shock-wave.
Originally develop to produce a 1-dimensional solution for a normal shock that is a bimodal mixture of two Maxwell--Boltzmann distributions parameterized
by the position along the streamline, the methodology has been recently
extended to 2-dimensional reflected shocks~\cite{timokhin2022mott}. However, we consider a simple 3-dimensional extension of the original bimodal formulation,
as the focus of the present work lies in the general problem of distribution reconstruction and not necessarily in the accurate representation of multi-dimensional
physical shock phenomena.
We assume the distribution to be given by
\begin{equation}
    f(\mathbf{v}) = \frac{\alpha}{C_{MS} n_1 T_1^{3/2}} \exp\left(-\frac{||\mathbf{v}-\mathbf{v}_1||_2^2}{T_1}\right)
    +              \frac{1-\alpha}{C_{MS} n_2 T_2^{3/2}} \exp\left(-\frac{||\mathbf{v}-\mathbf{v}_2||_2^2}{T_2}\right),\label{eq:mott-smith}
\end{equation}
where
\begin{equation}
    \alpha = \frac{1}{1 + \exp(x)},
\end{equation}
\begin{equation}
    n_1 = 1,\quad n_2 = M^2 \frac{\gamma + 1}{2 + M^2 (\gamma - 1)}
\end{equation}
\begin{equation}
    \mathbf{v}_1 = \left(\sqrt{\gamma} M, 0, 0\right),\quad \mathbf{v}_2 = \left(\frac{1}{n_2}\sqrt{\gamma} M, 0, 0\right),
\end{equation}
\begin{equation}
    T_1 = 1,\quad T_2 = \frac{1 - \gamma + 2 \gamma M^2}{1 + \gamma} \frac{1}{n_2}.
\end{equation}
Here $x$ is the position along the streamline and defines the mixing of the two Maxwell--Boltzmann distributions. The distribution parameters
with index 1 describe to the cold high-velocity gas before the shock at $x=-\infty$ moving at a speed with Mach number $M$, and values
with index 2 describes to the hot slow post-shock gas at $x=+\infty$. $\gamma$ is the ratio of specific heats, which we take as 5/3, corresponding to a monatomic gas.
The scaling $C_{MS}$ is used to scale the total number density to 1 for consistency with the other cases.
In the present work, we fix $x=0$ in order to obtain a bimodal distribution.

\begin{figure}[t!]
  \centering
  \includegraphics[width=0.85\textwidth]{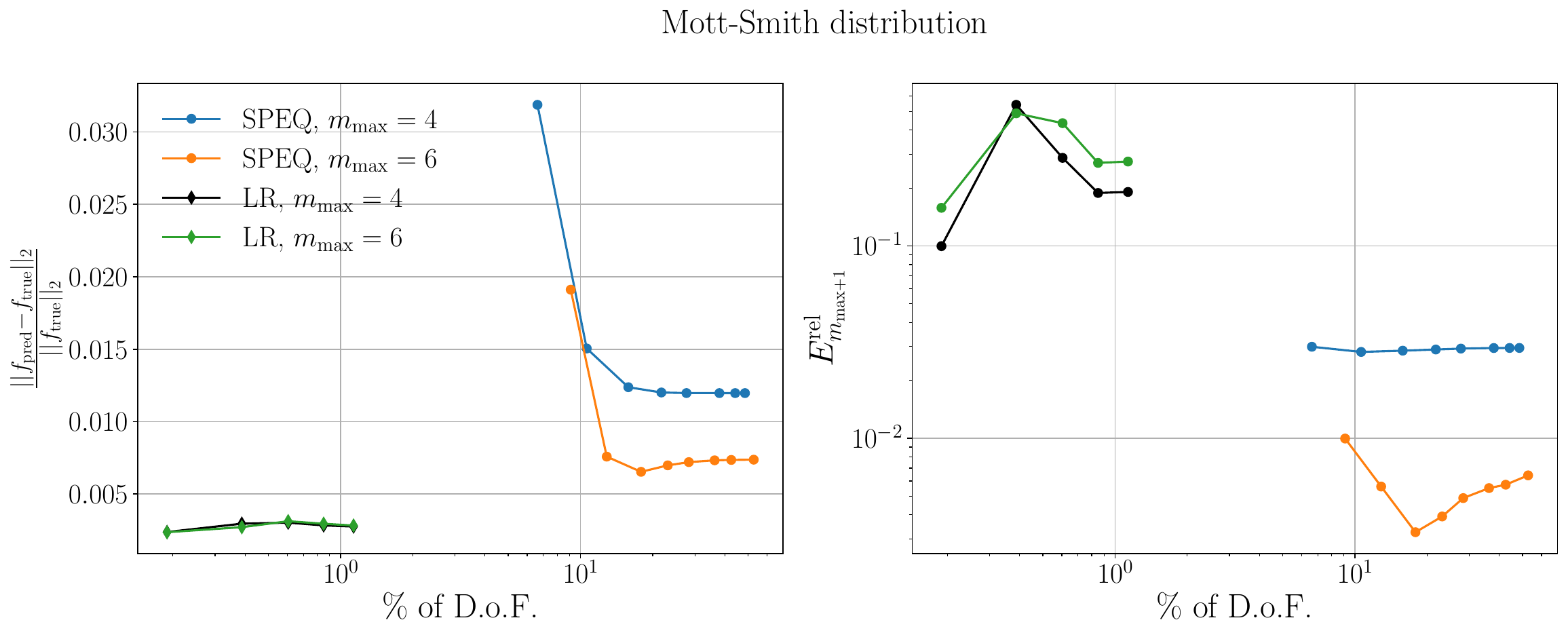}
  \caption{Relative $L_2$ error in the VDF (left) and relative error in next predicted moments (right) as a function
  of the fraction of used degrees of freedom, Mott-Smith distribution.}\label{fig:ms-errors}
  \end{figure}

\begin{figure}[t!]
  \centering
  \includegraphics[width=0.85\textwidth]{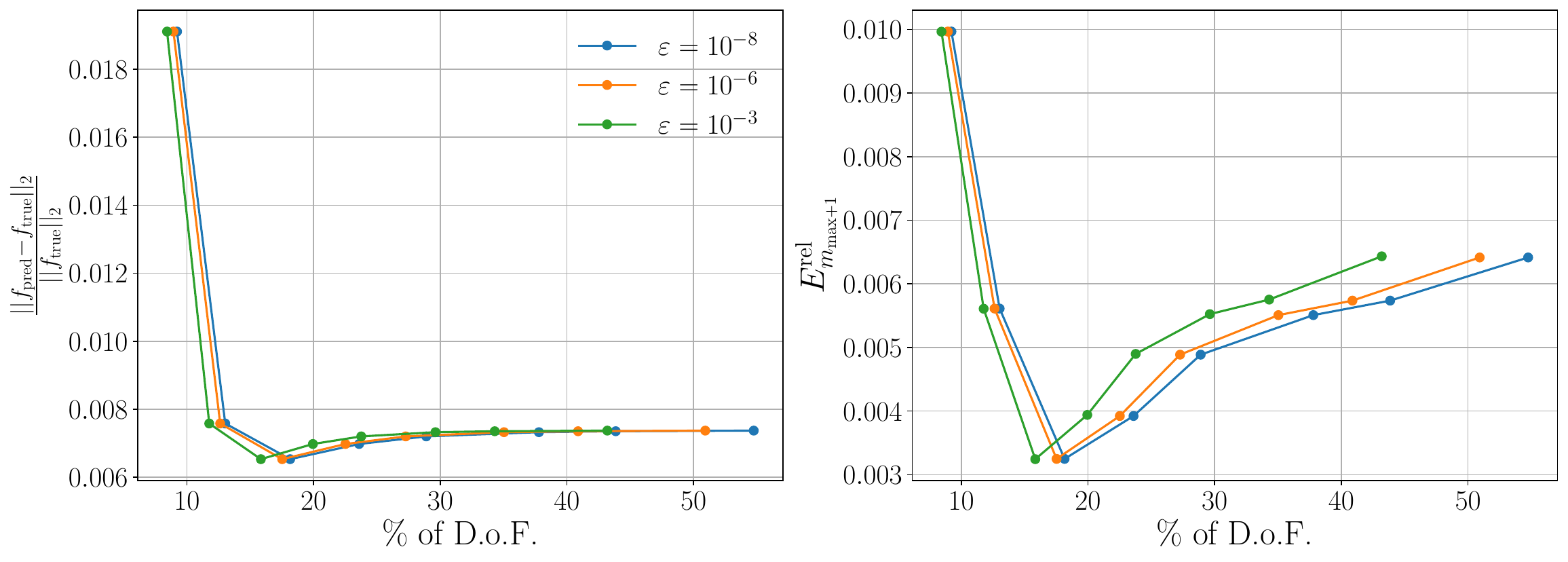}
  \caption{Relative $L_2$ error in the VDF (left) and relative error in next predicted moments (right) of total order $7$ as a function
  of the fraction of used degrees of freedom for different values of cut-off threshold parameter $\varepsilon$, Mott-Smith distribution.}\label{fig:ms-eps-study}
  \end{figure}

  \begin{figure}[t!]
  \centering
  \includegraphics[width=0.85\textwidth]{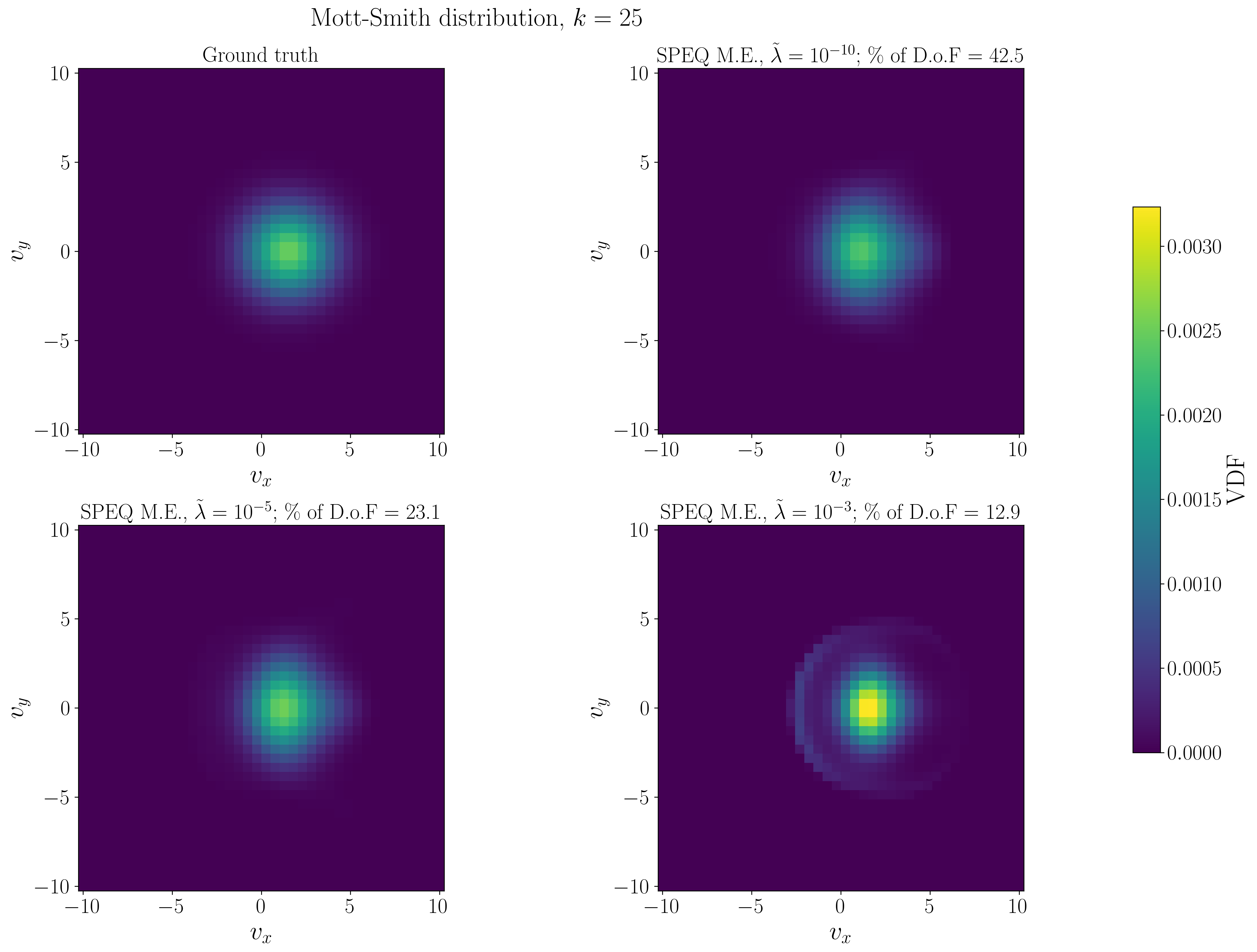}
  \caption{Slices along the $v_z$ axis ($k=25$) of the hidden truth distribution and reconstructed distributions for various values of $\lambda$, Mott-Smith distribution.}\label{fig:ms-vzslice}
  \end{figure}

  \begin{figure}[h!]
  \centering
  \includegraphics[width=0.85\textwidth]{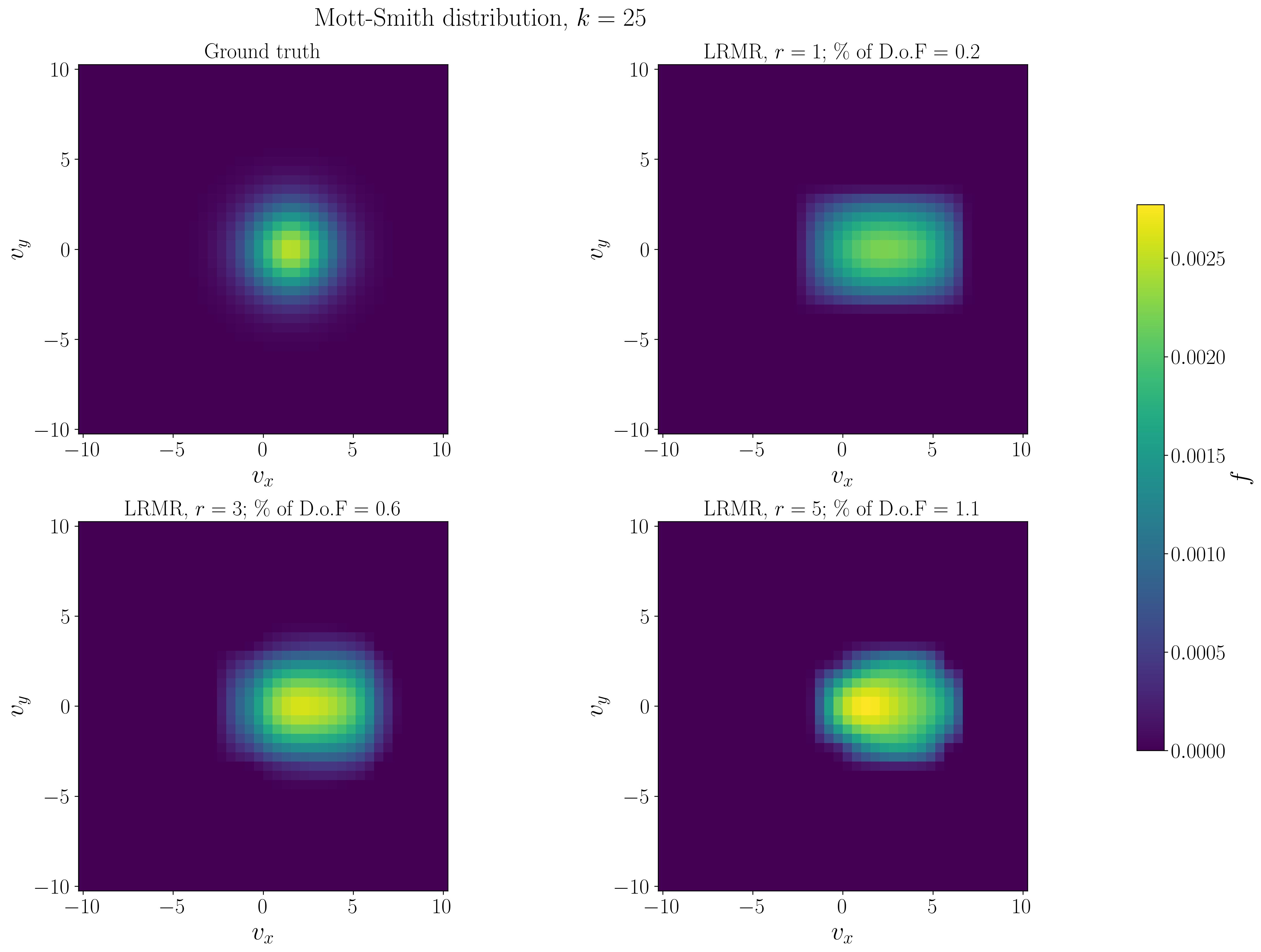}
  \caption{Slices along the $v_z$ axis ($k=25$) of the hidden truth distribution and reconstructed distributions for various values of $\lambda$, Mott-Smith distribution.}\label{fig:ms-lr-vzslice}
  \end{figure}

Figure~\ref{fig:ms-errors} shows the errors in the reconstructed distribution and the next predicted moments as a function of the fraction of used degrees of freedom.
For the SPEQ method, we see a plateau of the error in the reconstructed distribution at higher fractions of the degrees of freedom used, as the entropy minimization-based reconstruction is not
able to correctly reconstruct the distribution~\cite{yilmaz2024nonlinear}. Incorporating higher-order moments into the constraints leads to a reduction in both VDF reconstruction
error and error in the next predicted moments (as seen on the right subplot), consistent with the behaviour of the non-sparse entropy minimization-based approach~\cite{yilmaz2024nonlinear}.
It should be noted that even for high degrees of sparsity, the relative errors in the next predicted moments remain quite small, on the order of 1\%, and only a factor of 2 larger
 than the errors in the non-sparse solution.
 For the LRMR method, the role of the rank is not as clear, with the choice of $r$ not having a significant impact on the solution quality: the relative $L_2$ errors are noticeably lower than those of
 the SPEQ method, but the moment prediction methods are larger. Incorporation of a larger number of moment constraints also has a small impact on the solution quality. Since the problem is not well-posed,
 the method can locate a sub-optimal local minimum; therefore, we plan on investigating other objective functions that incorporate additional information apart from the moment constraints.

We also investigate the role the choice of the $\varepsilon$ parameter in the SPEQ method, which is used to determine which values of the predicted function $g$ are set to 0, plays in the reconstruction.
Figure~\ref{fig:ms-eps-study} shows the errors for different
values of $\varepsilon$. We see that the error in the reconstructed distribution is not very sensitive to the choice of $\varepsilon$, but at low values of $\lambda$,
higher values $\varepsilon$ lead to stronger sparsity; for larger values of $\lambda$, the $L_1$ regularization is strong enough that the exact value of the cutoff is less significant,
as the values of $g$ are already small enough.

Figure~\ref{fig:ms-vzslice} shows slices along the $v_z$ axis of the distribution function reconstructed using the SPEQ method for different values of $\lambda$.
We see that the reconstruction is able to capture the main features of the distribution function, such as the bimodal structure; very strong $L_1$ regularization however leads
to noticeable artifacts in the distribution. However, less sparse reconstructions do not exhibit them. Figure~\ref{fig:ms-lr-vzslice} shows the same slices of the distributions
obtained with the LRMR method, which at higher ranks does resemble the underlying distribution and exhibits less artifacts than the SPEQ results, even though an order of magnitude fewer
degrees of freedom are required.

\subsection{Collisionless plasmas}\label{sec:collisionlessplasmas}

\begin{figure}[t!]
\centering
\includegraphics[width=0.85\textwidth]{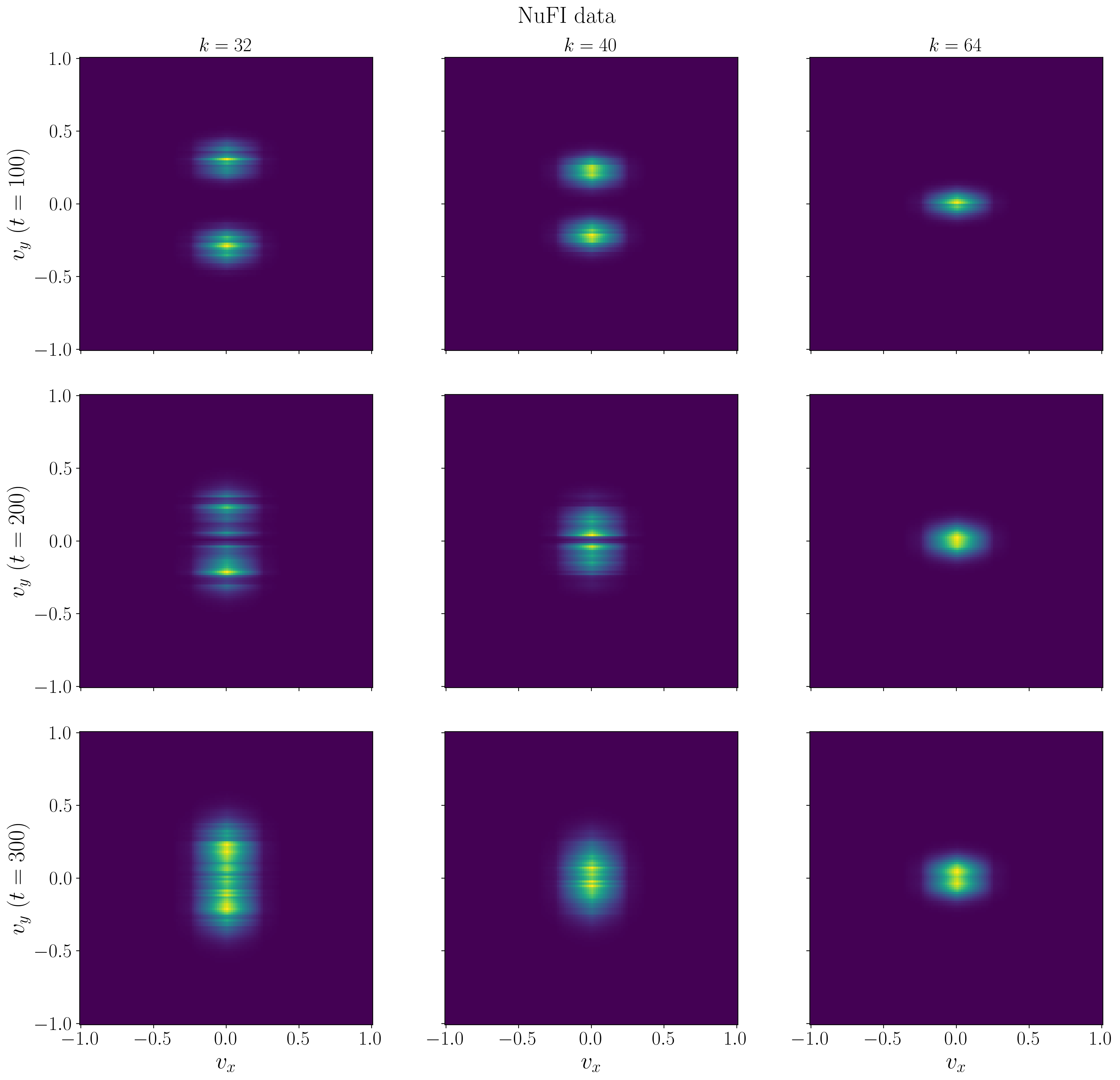}
\caption{Results of the beam-driven instability simulation produced by NuFI. Different rows correspond to different times, and different columns correspond to different $v_z$-slices.}\label{fig:nufi-data}
\end{figure}

Finally, we look at velocity distribution functions arising as the solution of
the collisionless Vlasov--Maxwell equation, i.e., equation~\eqref{eqn:kinetic}
with $Q = 0$ and electro-magnetic forces $F = \tfrac{q}{m}(E + v \times B)$.
The Vlasov--Maxwell equation is coupled to the Maxwell equations
\begin{gather}
\label{eqn:Maxwell_1}
\nabla \times E + \partial_t B = 0, \qquad
\nabla \times B - \partial_t E = j, \\
\label{eqn:Maxwell_2}
\nabla \cdot E =  \rho, \qquad
\nabla \cdot B = 0,
\end{gather}

where the right hand sides are computed from the distribution function f via
\begin{gather}
\label{eqn:rho_j}
\rho(t,x) = \int_{\mathbb{R}^d} f(t,x,v) \text{dv}, \qquad
j(t,x) =  \int_{\mathbb{R}^d} v f(t,x,v) \text{dv}.
\end{gather}

Collisionless plasmas are commonly encountered in e.g. astrophysical settings
such as the solar wind. Due to the lack of collisions these plasmas tend to
locally develop very fine scale structures, called filamentation, which poses
significant numerical challenges.
Resolving this filamentation is crucial to capture the onset of kinetic
instabilities or other relevant kinetic effects such as Landau Damping, which
govern the energy transfer between the scales.

The simulation to generate the present data is performed using the Numerical
Flow Iteration (NuFI), which is based upon the Hamiltonian splitting of the
Vlasov--Maxwell system.~\cite{wilhelmBacchini2025, kirchhart2023numerical, wilhelm2025extendingnumericalflowiteration}
Directly approximating the characteristic flow by splitting the Hamiltonian
sub-flows, allows resolving fine scales far below the grid scale (called the
Zoom property), while also preserving the conservation properties of the
analytical solution.

Here we consider the simulation of a beam-driven
instability:~\cite{wilhelmBacchini2025}
On the time scales we are interested in, it is safe to assume a uniform ion
background. The initial electron distribution function is
\begin{equation}
    \footnotesize
f_0(x,y,u,v,w) = \tfrac{1 + \alpha \cos(x) \cos(y)}{2(2 \pi
v_{\text{th}}^2)^{3/2}}
\left(\exp\left(-\tfrac{(u - v_d)^2}{2v_{\text{th}}^2}\right)
+ \exp\left(-\tfrac{(u + v_d)^2}{2v_{\text{th}}^2}\right)\right)
\exp\left(-\tfrac{v^2 + w^2}{2 v_{\text{th}}^2}\right),
\end{equation}
where we set $v_{\text{th}} = 1$, $\alpha = 0.01$ and $v_d = 2$. The initial
electric field can be computed from the Gauss law as
$E_0(x,y) = \tfrac{\alpha}{2}(\sin(x)\cos(y), \cos(x) \sin(y), 0)^T$.
To trigger magnetic effects we also impose an initial magnetic field
$B_0(x,y) = B_0 (0, 0, \cos(x)\cos(y))^T$ with $B_0 = 0.1$, which in the late
stage of the simulations triggers the formation of magnetic islands.
The configuration space is set to $[0,2\pi] \times [0,2\pi]$
and the numerical velocity space is chosen as $[-30,30] \times [-30,30]
\times [-10,10]$. The field interpolation is using B-Splines of 4th order.
To compute the current density we use $16 \times 16$ points in space and
$48 \times 48 \times 16$ in velocity.
Note that in contrast to classical grid-based (or particle-based) methods this
choice of phase-space discretization does not restrict us in the plotting
resolution. While it affects the discretization errors, NuFI still allows us
to zoom arbitrarily far into any region of phase-space when plotting, which we
use to generate high resolution plots of velocity distribution functions.

\begin{figure}[t!]
\centering
\includegraphics[width=0.85\textwidth]{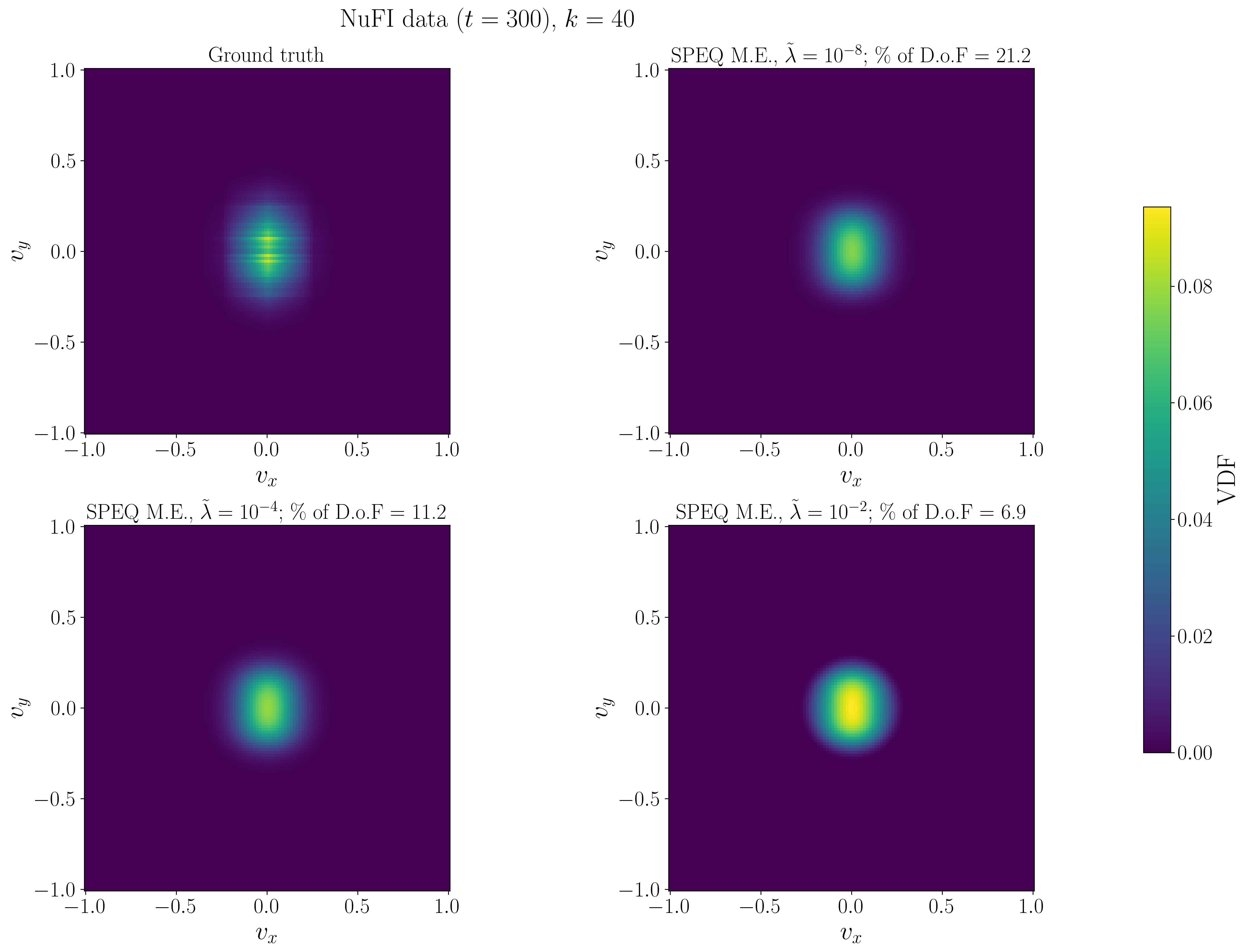}
\caption{Slices along the $v_z$ axis ($k=40$) of the hidden truth distribution and distributions reconstructed using SPEQ with entropy minimization for various values of $\lambda$, beam-driven instability at $t=300$; reference NuFI solution in the upper left.}\label{fig:nufi-me-vz40}
\end{figure}

The initial electro-magnetic perturbation trigger a beam-driven instability,
which manifests through a linear growth in the electric energy between
$t \approx 20 $ and $120$ at which point it saturates and the non-linear phase
of the instability is reached. The magnetic energy is initially slightly
dampened until about $t \approx 100$ after which it experiences a short linear
growth phase to finally reach a higher saturation level at the same time as the
electric field.
The phase space distribution functions, which interest us in this work develop
increasingly fine scale structure such as small-scale vortices (corresponding
to particle trapping) and magnetic islands.

Since in this test case, the underlying distribution is known,
and the goal is a storage-efficient approximation thereof, the velocity grid can be re-scaled arbitrarily, as the moments are computed
based on the known underlying distribution. Therefore, we scale the grid to a $[-1,1] \times [-1,1]
\times [-1,1]$ one, to improve the conditioning of the moment measurement matrix.

For the SPEQ approach, we take the following $\lambda$ values:
\begin{equation}
    \tilde{\lambda} := \frac{\lambda}{\Delta v} \in \{0, 10^{-8}, 10^{-6}, 10^{-4}, 10^{-2}, 1, 10, 100\},
\end{equation}
and a threshold $\varepsilon=10^{-7}$ is used to cut off the values of $g$.

Figure~\ref{fig:nufi-data} shows the simulation
results for various times for different $v_z$-slices, as given by the index $k$. The data are
saved on a $128^{3}$ velocity grid; values of $k$ closest to 64 correspond to the slices in the bulk of the distribution
with a small $v_z$ velocity, whilst values of $k$ closest to 0 and 128 correspond to the slices with the largest absolute $v_z$ velocity.

Whilst the maximum entropy method is capable of representing a large variety of distributions~\cite{boccelli2024gallery}, it struggles with
non-smooth distributions and sharp distribution gradients --- exactly the types of distributions produced by NuFI, which can capture
very fine filamentations in the plasma. As such, it is expected that the maximum entropy approach will exhibit noticeable errors
when applied to the beam instability results. However, since in this case the underlying distribution is known, and we aim to simply
compress it whilst retaining key features, we can also make use of the generalized formulation~(\ref{eq:optimization-spec-KL})
and minimize not the entropy, but the Kullback--Leibler divergence to the distribution values obtained from NuFI.
Thus, we consider the following two approaches:
\begin{enumerate}
    \item the sparse entropic quadrature method~(\ref{eq:optimization-spec}) using the estimated local Maxwellian as the weighting function $w$
    \item the generalized formulation~(\ref{eq:optimization-spec-KL}) with both $\phi_{ijk}$ and the weighting function being the NuFI data, i.e.\ $w_{ijk}=\phi_{ijk}$
\end{enumerate}

\begin{figure}[t!]
\centering
\includegraphics[width=0.85\textwidth]{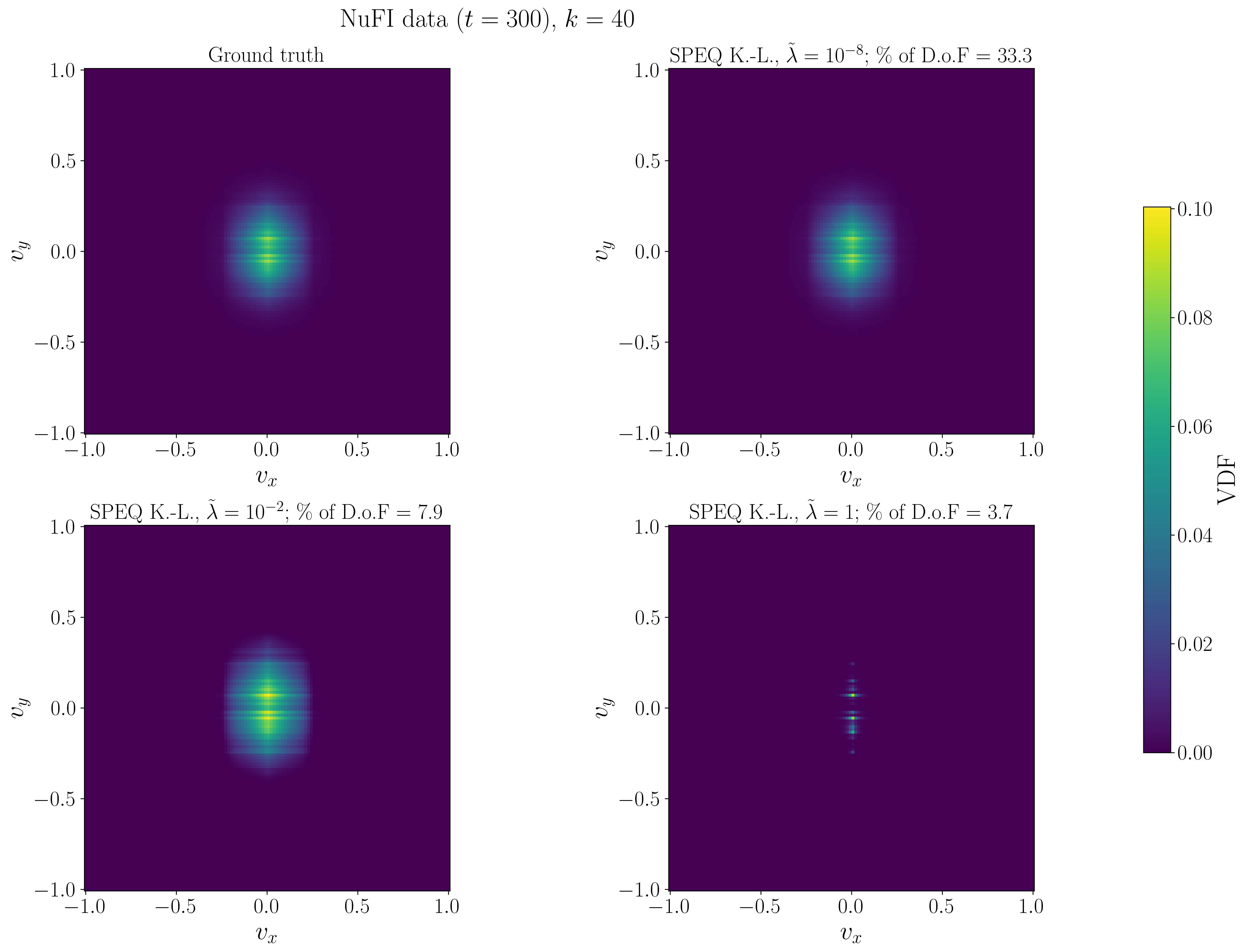}
\caption{Slices along the $v_z$ axis ($k=40$) of the hidden truth distribution and distributions reconstructed using SPEQ with minimization of Kullback-Leibler divergence w.r.t the original data for various values of $\lambda$, beam-driven instability at $t=300$; reference NuFI solution in the upper left.}\label{fig:nufi-kl-vz40}
\end{figure}

\begin{figure}[t!]
\centering
\includegraphics[width=0.85\textwidth]{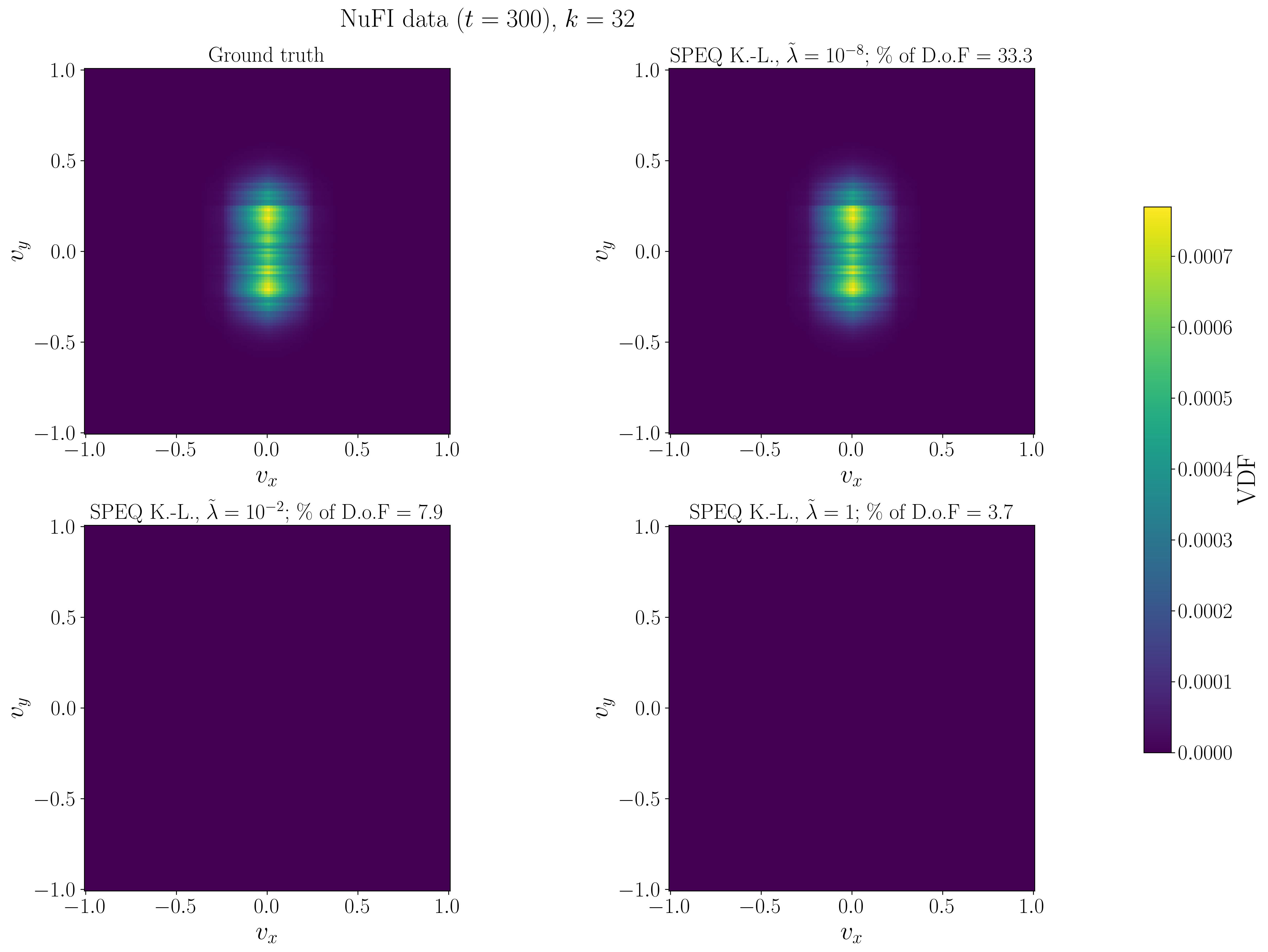}
\caption{Slices along the $v_z$ axis ($k=32$) of the hidden truth distribution and distributions reconstructed using SPEQ minimization of Kullback-Leibler divergence w.r.t the original data for various values of $\lambda$, beam-driven instability at $t=300$; reference NuFI solution in the upper left.}\label{fig:nufi-kl-vz32}
\end{figure}

\begin{figure}[t!]
\centering
\includegraphics[width=0.85\textwidth]{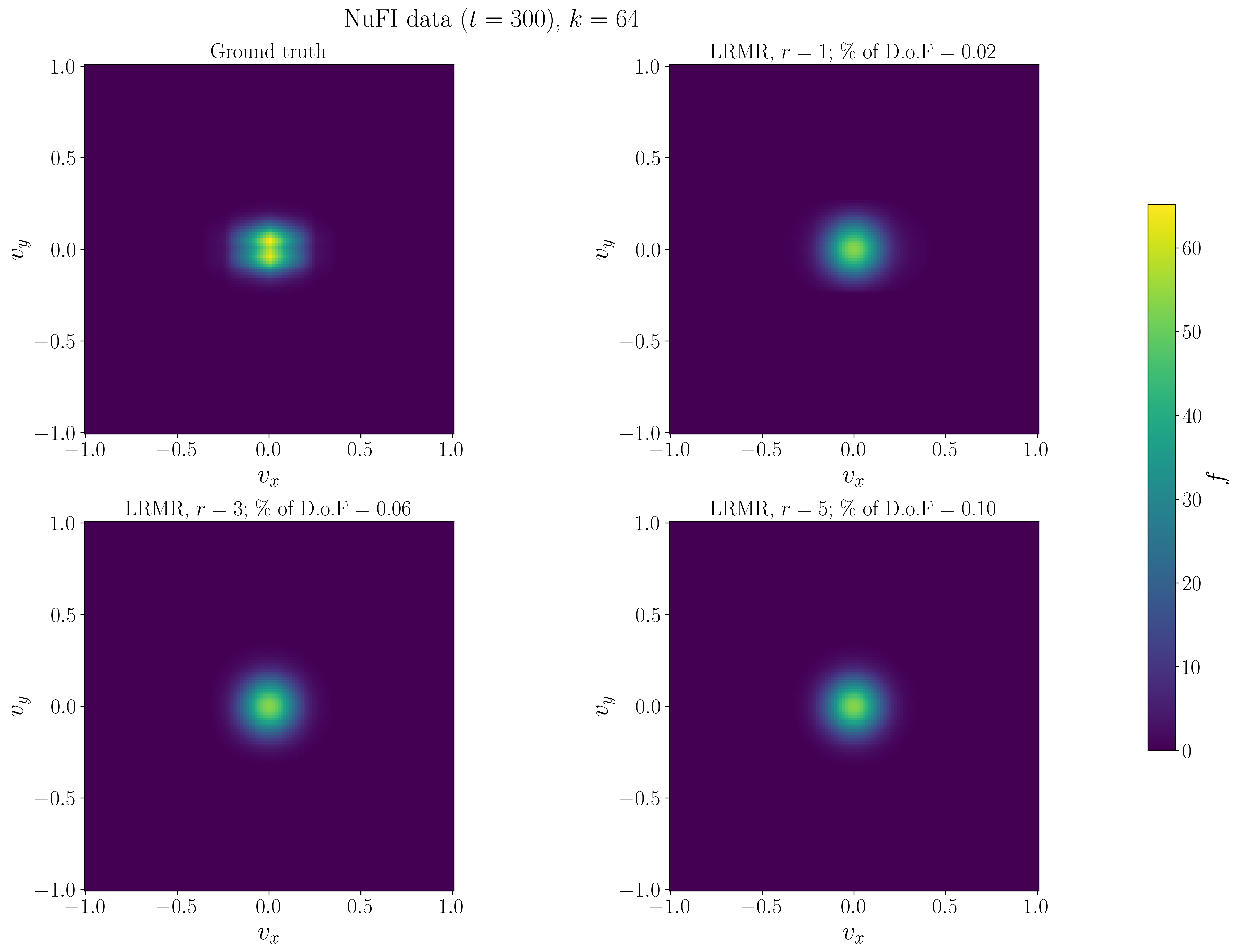}
\caption{Slices along the $v_z$ axis ($k=64$) of the hidden truth distribution and distributions reconstructed using the LRMR method for various ranks $r$, beam-driven instability at $t=300$; reference NuFI solution in the upper left.}\label{fig:nufi-kl-lrmr-vz64}
\end{figure}

\begin{figure}[t!]
\centering
\includegraphics[width=0.85\textwidth]{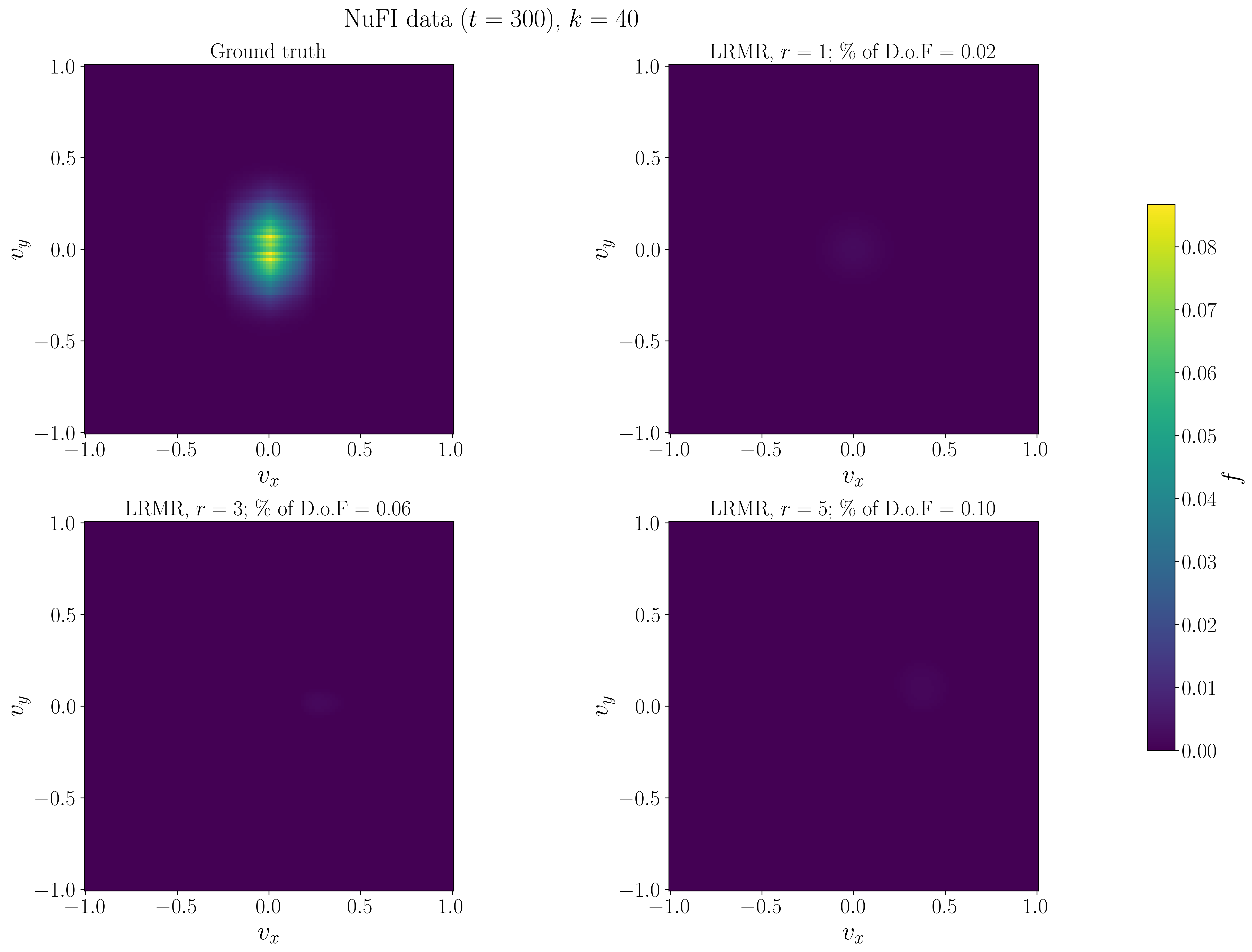}
\caption{Slices along the $v_z$ axis ($k=40$) of the hidden truth distribution and distributions reconstructed using the LRMR method for various ranks $r$, beam-driven instability at $t=300$; reference NuFI solution in the upper left.}\label{fig:nufi-kl-lrmr-vz40}
\end{figure}

Figure~\ref{fig:nufi-me-vz40} shows a slice of the distribution function as predicted by the sparse entropic quadrature method, which produces
a significantly smoother distribution than the reference data shown in the upper-left subplot, only retaining the very general shape of the
velocity distribution function and losing physically significant details.

Therefore, we turn to the generalized SPEQ approach minimizing the Kullback-Leibler divergence. Figure~\ref{fig:nufi-kl-vz40} shows
the sparse distributions obtained with this method for the same values of $\lambda$ and the same slice as shown on Figure~\ref{fig:nufi-me-vz40}.
We observe a significantly better representation of the distribution, with only very sparse solutions differing noticeably
from the reference data, as seen in the bottom-right subplot, although they still retain some of the filamented structure.

Looking at slices of the distribution function further out in the tails, as shown on Figure~\ref{fig:nufi-kl-vz32}, we see that sparser solutions
produce zero-valued distributions in this region, however, less sparse reconstructions retain the expected features. It should be noted that the values
as seen on Figure~\ref{fig:nufi-kl-vz32} are two orders of magnitude smaller than the values shown on Figure~\ref{fig:nufi-kl-vz40}, and it is therefore
expected that sparsification will lead to loss of detail in the tails of the distribution.

For the low-rank distributions recovered with the LRMR method, we plot slightly different slices, namely at $k=64$ and $k=40$, as shown on Figures~\ref{fig:nufi-kl-lrmr-vz64}--\ref{fig:nufi-kl-lrmr-vz40}.
On Figure~\ref{fig:nufi-kl-lrmr-vz64} we observe that already in the middle of the distribution LRMR recovers a much more uniform distribution which decays significantly faster in the tails
than the actual underlying data, as evidenced by the much smaller values of the recovered distributions on Figure~\ref{fig:nufi-kl-lrmr-vz40}.

\begin{figure}[h!]
\centering
\includegraphics[width=0.85\textwidth]{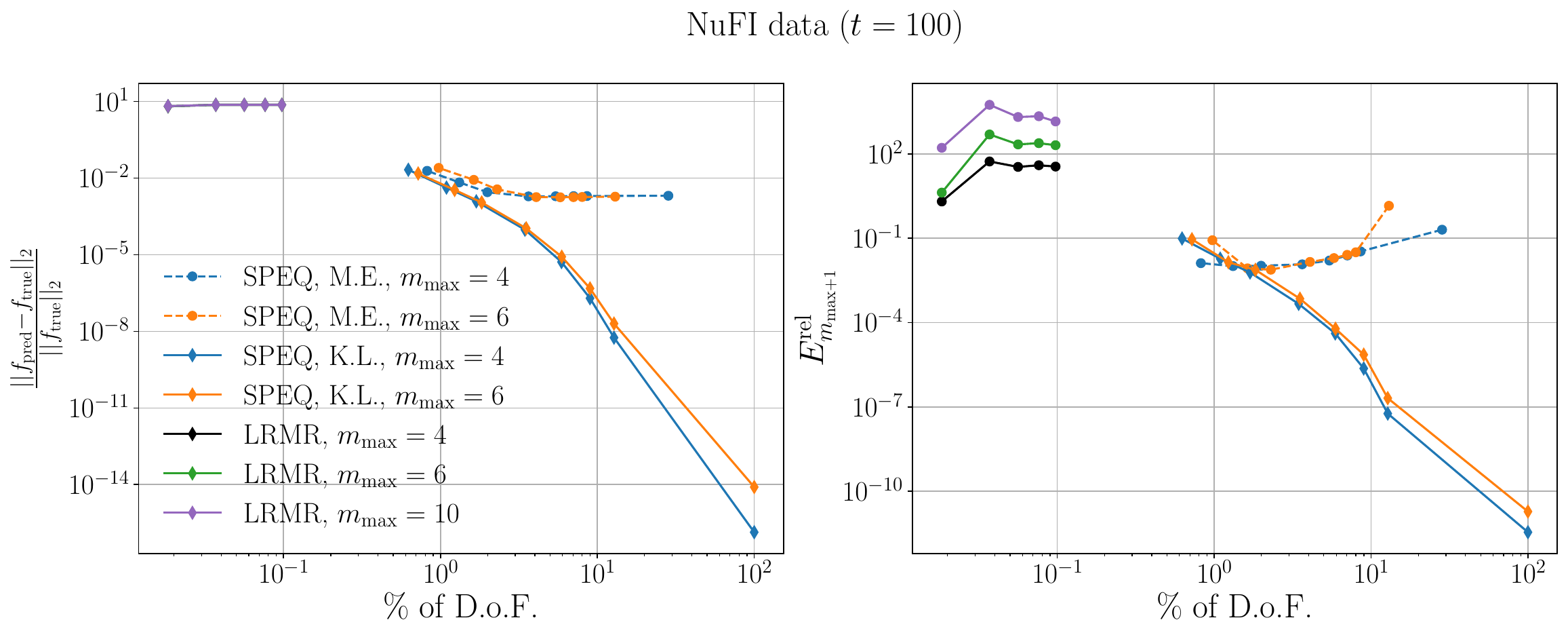}
\caption{Relative $L_2$ error in the VDF (left) and absolute error in next predicted moments (right) as a function
  of the fraction of used degrees of freedom, NuFI simulation results at $t=100$.}\label{fig:nufi-errors_t100}
\end{figure}

\begin{figure}[h!]
\centering
\includegraphics[width=0.85\textwidth]{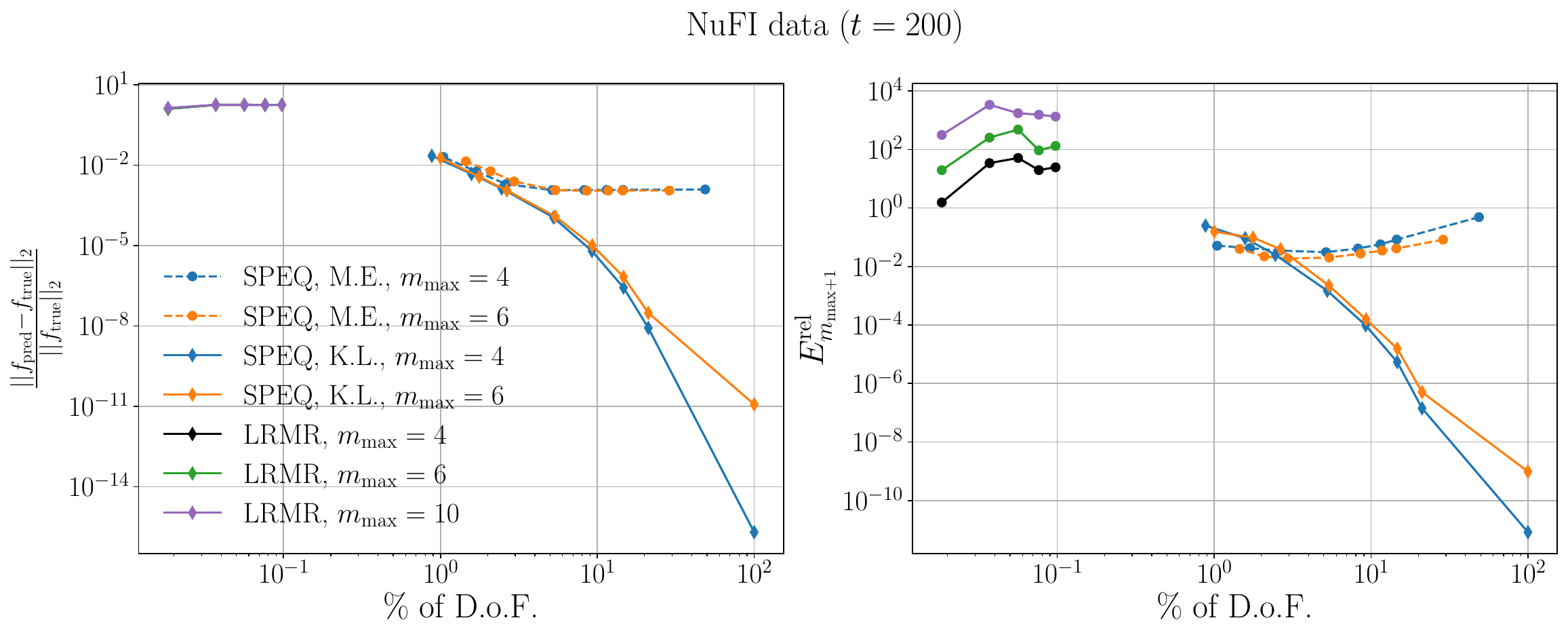}
\caption{Relative $L_2$ error in the VDF (left) and absolute error in next predicted moments (right) as a function
  of the fraction of used degrees of freedom, NuFI simulation results at $t=200$.}\label{fig:nufi-errors_t200}
\end{figure}

\begin{figure}[h!]
\centering
\includegraphics[width=0.85\textwidth]{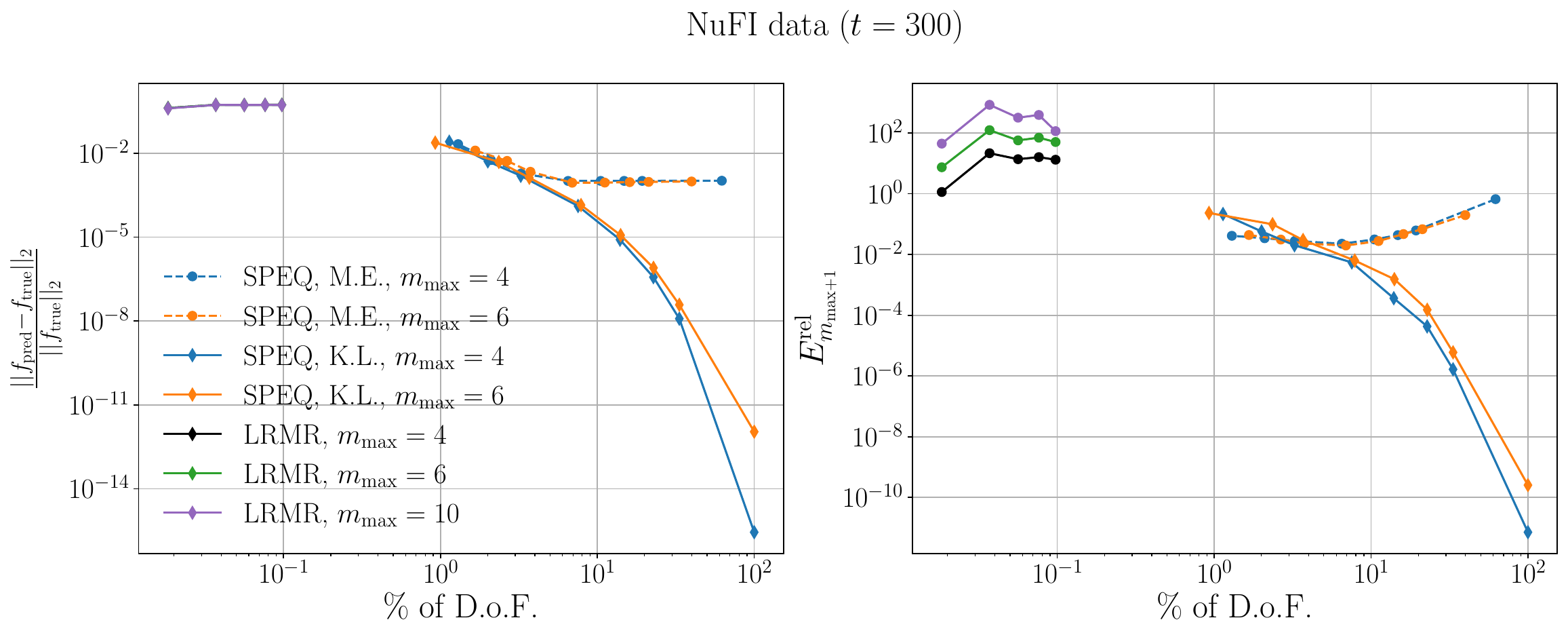}
\caption{Relative $L_2$ error in the VDF (left) and absolute error in next predicted moments (right) as a function
  of the fraction of used degrees of freedom, NuFI simulation results at $t=300$.}\label{fig:nufi-errors_t300}
\end{figure}

Figures~\ref{fig:nufi-errors_t100}--\ref{fig:nufi-errors_t300} show the relative $L_2$ errors in the reconstructed distribution and
the relative errors in the predicted moments as a function of the number of degrees of freedom used to represent the distribution for
reconstructions of the beam instability data at $t=100$, $t=200$, and $t=300$, respectively. ``SPEQ M.E.'' denotes solutions obtained
via solving the entropy minimization problem, whereas ``SPEQ K.L.'' denotes solutions computed via minimizing the Kullback--Leibler divergence.
For the LRMR approach, we additionally consider conservation of all moments up to order 10, as in theory, given that we know the underlying distribution, we can construct moment measurement
matrices of arbitrary order, although this leads to significant usage of machine memory and higher computational costs.
Due to the finer structure appearing at later times, as seen on Figure~\ref{fig:nufi-data}, the Kullback--Leibler-based reconstruction exhibits higher errors for later times (Figure~\ref{fig:nufi-errors_t300}), but in general behaves as expected --- less sparsity leads to more accurate results.
The entropy minimization-based approach
however leads to plateauing errors even for low values of $\lambda$, as it effectively reconstructs a smooth distribution satisfying the moment constraints, which
is related to the reference data only through these constraints. In fact, introducing a certain degree of sparsity actually leads to a slight improvement
in the moment prediction error.
We also note that both approaches produce higher degrees of sparsity for earlier times, compare Figure~\ref{fig:nufi-errors_t100} to Figure~\ref{fig:nufi-errors_t300}, as less finer detail is present in the original distribution. The errors of the LRMR method are virtually unaffected by the number of moments used as constraints; thus, additional data- or physics-informed priors are needed to achieve
better solution accuracy.

For the SPEQ method, conservation of a larger number of moments also has little impact on the solution quality. For the entropy minimization-based approach, the errors are in
general large, and it is likely that incorporation of additional moment information changes the predicted distribution insufficiently to have a noticeable effect on the errors. In the approach minimizing the Kullback--Leibler divergence to the reference data, the optimization target is
the main influencer of the solution quality.

We can conclude that for high-resolution data produced by kinetic solvers, the Kullback--Leibler-based reconstruction using SPEQ can produce reasonable approximations of the underlying data even with high degrees of sparsity, whereas the entropy minimization-based and LRMR approaches lead to an almost
complete loss of the fine detail.

\section{Conclusions}\label{sec:conclusions}
We propose an extension of the entropic quadrature method that allows for enforcement of sparsity in the resulting representation of the velocity distribution function and can be also used to produce sparse representations of known distribution data. We also propose a low rank-constrained moment recovery method that imposes a low-rank structure on the reconstructed distribution whilst aiming to preserve
a given set of moments.
The approaches are tested against several gas dynamics distributions, both analytical and high-resolution numerical results of collisionless plasma
simulations. The reconstruction methods can achieve high degrees of sparsity whilst still retaining low error in the predicted distributions and
moments thereof, although for highly irregular distributions with fine details, such as those obtained in high-resolution collisionless plasmas simulations,
the method struggle to retain these structures.
Future work will focus on the application of the reconstruction-based moment closure methods to dynamical problems and the use of the sparse and low-rank reconstructions
as restarts for the NuFI solver. Further, we will investigate the use of other target objectives~\cite{abdelmalik2016moment,ghalamkari2026deformed} for the SPEQ and LRMR methods, respectively, and focus on improving
the computational performance of the reconstruction methods.

\section{Acknowledgments}
GO, LT, MH, and MT thank the Deutsche Forschungsgemeinschaft (DFG, German Research Foundation)
for financial support through the SFB1481 ``Sparsity and Singular Structures'' (442047500) within the project B04 ``Sparsity Patterns in Kinetic Theory''.
R.P.W. thanks the European High Performance Computing Joint Undertaking (JU) and Belgium, Czech Republic, France, Germany, Greece, Italy, Norway, and Spain
for supporting this work under grant agreement No 101093441. Views and opinions expressed are however
those of the author(s) only and do not necessarily reflect those of the
European Union or the European High Performance Computing Joint Undertaking
(JU) and Belgium, Czech Republic, France, Germany, Greece, Italy, Norway, and
Spain. Neither the European Union nor the granting authority can be held
responsible for them.






\end{document}